\RequirePackage{ifpdf}
\documentclass[12pt]{article}
\ifpdf
	\usepackage[pdftex]{graphicx,color}
\else
	\usepackage[dvips]{graphicx}
\fi
\usepackage{epsfig}
\usepackage{latexsym,amsmath,amsfonts,amssymb}
\usepackage[latin1]{inputenc}
\usepackage[american]{babel}
\usepackage{hyperref}
\usepackage{bbm}
\newcommand{\beq}{\begin{equation}}
\newcommand{\eeq}{\end{equation}}
\newcommand{\beqn}{\begin{equation*}}
\newcommand{\eeqn}{\end{equation*}}
\newcommand{\bear}{\begin{eqnarray}}
\newcommand{\eear}{\end{eqnarray}}
\newcommand{\bal}{\begin{aligned}}
\newcommand{\eal}{\end{aligned}}
\newcommand{\nn}{\nonumber}

\begin{document}
\baselineskip=15.5pt
\pagestyle{plain}
\setcounter{page}{1}

\newcommand{\cN}{{\cal N}}
\newcommand{\cO}{{\cal O}}
\newcommand{\cA}{{\cal A}}
\newcommand{\cT}{{\cal T}}
\newcommand{\cF}{{\cal F}}
\newcommand{\cC}{{\cal C}}
\newcommand{\cR}{{\cal R}}
\newcommand{\cW}{{\cal W}}
\newcommand{\cK}{{\cal K}}



\def\a{\alpha}
\def\b{\beta}
\def\c{\gamma}
\def\d{\delta}
\def\eps{\epsilon}           
\def\f{\phi}               
\def\vf{\varphi}  \def\tvf{\tilde{\varphi}}
\def\vp{\varphi}
\def\g{\gamma}
\def\h{\eta}
\def\j{\psi}		\def\y{\psi}
\def\k{\kappa}                    
\def\l{\lambda}
\def\m{\mu}
\def\n{\nu}
\def\o{\omega}  \def\w{\omega}
\def\p{\pi}
\def\q{\theta}  \def\th{\theta}                  
\def\r{\rho}                                     
\def\s{\sigma}                                   
\def\t{\tau}
\def\u{\upsilon}
\def\x{\xi}
\def\z{\zeta}
\def\pt{\tilde{\varphi}}
\def\tt{\tilde{\theta}}
\def\lab{\label}
\def\6{\partial}
\def\wg{\wedge}
\def\atanh{{\rm arctanh}}
\def\bpsi{\bar{\psi}}
\def\bt{\bar{\theta}}
\def\bvf{\bar{\varphi}}
\def\W{\Omega}

\numberwithin{equation}{section}
\newcommand{\Tr}{\mbox{Tr}}    
\renewcommand{\theequation}{{\rm\thesection.\arabic{equation}}}
\allowdisplaybreaks[1]


\newcommand{\uw}{\underline{\w}}
\newcommand{\us}{\underline{\s}}
\newcommand{\tN}{\tilde{N}}
\newcommand{\hN}{\hat{N}}
\newcommand{\mA}{\mathcal{A}}
\newcommand{\mB}{\mathcal{B}}
\newcommand{\cH}{\mathcal{H}}
\newcommand{\mS}{\mathbb{S}}
\newcommand{\cS}{\mathcal{S}}
\newcommand{\mH}{\mathbb{H}}
\newcommand{\mR}{\mathbb{R}}
\newcommand{\td}{\mathrm{d}}
\newcommand{\vol}[1]{\textrm{Vol(}#1\textrm{)}}
\renewcommand{\arraystretch}{1.5}
\newcommand{\SxS}{\mS^2\times\mS^3}
\newcommand{\SL}{\widetilde{SL_2}}
\newcommand{\HxH}{\mH_2\times\SL}
\newcommand{\pbi}[1]{\imath^*\left(#1\right)}


\begin{titlepage}
\vspace{0.1in}

\begin{center}
\Large \bf Kutasov-like duality from D5-branes wrapping hyperbolic cycles 
\end{center}
\vskip 0.2truein
\begin{center}
Eduardo Conde  ${}^{\dagger}$\footnote{eduardo@fpaxp1.usc.es} and
J\'er\^ome Gaillard ${}^{\dagger \ddagger}$\footnote{pyjg@swansea.ac.uk}\\
\vspace{0.2in}
${}^{\dagger}$\it{
Departamento de  F\'{\i}sica de Part\'{\i}culas, Universidade
de Santiago de
Compostela\\and\\Instituto Galego de F\'{\i}sica de Altas
Enerx\'{\i}as (IGFAE)\\E-15782, Santiago de Compostela, Spain
}
\\
\vspace{0.2in}
${}^{\ddagger}$ 
{\it Department of Physics, Swansea University \\ Singleton Park, Swansea SA2 8PP, United Kingdom
}

\vspace{0.2in}
\end{center}
\vspace{0.2in}
\begin{center}
	{\bf Abstract}
\end{center}

We study the addition of $N_f$ flavor D5-branes to supergravity solutions describing D5-branes wrapping two-cycles of genus $g>1$ inside a six-dimensional space equipped with an $SU(3)$-structure. The non-zero genus $g$ on the gravity side is dual to the existence of massless adjoint chiral superfields. Three types of internal manifolds are considered, each involving one of the following fibered products: $\HxH$, $\mS^2 \times \SL$ or $\mH_2 \times \mS^3$, where $\SL$ stands for the universal cover of $SL(2,\mR)$. For the first one, we investigate the dual field theories. We show that some of the solutions with $N_f \neq 0$ are dual to four-dimensional $\mathcal{N} = 1$ field theories exhibiting a Kutasov-like duality taking $N_c \rightarrow k  N_f - N_c$ and keeping $N_f$ fixed. Computed from the supergravity picture, $k$ is in general a rational number, which can be made integer to fit the expectation from the field theory side. We finally study some other properties of those field theories.

\smallskip
\end{titlepage}
\setcounter{footnote}{0}

\tableofcontents

\section{Introduction}

The AdS/CFT correspondence \cite{Maldacena:1997re,others} has been a quite active field from its very inception. Since then, it has greatly evolved, and it now stands as the best understood example of the general belief that many quantum field theories, at least in a certain regime where they are strongly coupled, should admit a description in terms of a string theory in higher dimensions. When the QFT is a gauge theory in its planar limit, we call this field of study the gauge/gravity correspondence: it establishes the dual character between a given gauge theory in its large `t Hooft coupling regime and a corresponding weakly coupled string theory, that reduces in this correspondence to a gravity theory.

Although the long-standing goal of finding a dual to QCD has not been achieved yet, the correspondence (which has not been formally proven), has been thoroughly tested through many examples, and we have learned many exciting things about the way it works. A lot of these examples beautifully show how certain QFT-like features can be realized geometrically. But a full understanding of the gauge/gravity duality is still lacking, and with each new example, new details are unveiled. In this paper, we would like to contribute one more example, which we hope provides some new insight.

We will be looking at supergravity solutions corresponding to branes wrapping compact cycles, an approach set forth by \cite{Maldacena:2000mw}. Wrapping branes on cycles of non-trivial homology does not seem to be a much explored avenue (see however \cite{Maldacena:2000mw,Paredes:2004xw,hyperbolic}), although a lot of mathematical structure appears, that relates to interesting physics. One recent example is the construction by Gaiotto and Maldacena \cite{Gaiotto:2009gz}, using M5-branes wrapping Riemann surfaces of arbitrary genus, of the gravity duals of certain $\cN=2$ super-conformal theories previously found by Gaiotto \cite{Gaiotto:2009we}. We are interested here in a related configuration, yielding different physics though: we will wrap D5-branes on Riemann surfaces with genus $g>1$, preserving only four supercharges. We refer to such surfaces as {\it hyperbolic cycles} since we build them as quotients of hyperbolic spaces. Our main motivation for doing this is to look for the gravity duals of theories displaying Kutasov duality \cite{Kutasov}, that appears when one has a non-trivial adjoint matter content in an $SU(N_c)$ gauge theory with fundamental matter. The fact that the adjoint content is non-trivial is directly related to having $g>1$. One formal way to explain this is using the index theorem like in \cite{Maldacena:2000mw}, that determines the number of fermion zero-modes from the topology of the space. As shown there, having a non-trivial genus $g>1$ implies the existence of $(g-1)$ massless adjoint fermions. Another way to think about those adjoints is that they roughly correspond to the zero-modes of the $B$-field on the cycles of different homology within the Riemann surface.

As we said, Kutasov duality involves the presence of fundamental matter in the gauge theory. The way to implement this on the gravity side is to introduce a smeared distribution of branes. This technique was first used in \cite{Bigazzi:2005md,Casero:2006pt}. For a recent review on how to use it for the addition of unquenched flavor in the gauge/gravity correspondence, one can have a look at \cite{Nunez:2010sf}. The basic idea is that introducing a set of branes (called flavor branes) in a supergravity solution is dual to introducing fundamental degrees of freedom (flavors) in the gauge theory. In order to explore the Veneziano limit of this gauge theory, one has to take the number of flavors $N_f\sim N_c$.  This implies that $N_f$ will be large as well as $N_c$, meaning for the dual supergravity solution that one cannot neglect the backreaction of the flavor branes on the geometry. The modified action one has to study is then
\beq
	S = S_{IIB} + S_{\text{sources}}\,,
	\label{eqn:S}
\eeq
where $S_{\text{sources}}$ will be the sum of the actions of the $N_f$ flavor branes. If these branes are coincident, the resulting system will be generally breaking some of the isometries (which are global symmetries of the field theory according to the AdS/CFT dictionary) of the original unflavored background. On top of that, the equations of motion of such a configuration will form a system of coupled non-linear PDE's, in general very hard to solve. The smearing technique solves both of these problems by exploiting the fact that there are a large number of flavor branes: one smears their distribution over some of the coordinates transverse to them, restoring the original isometries, and yielding equations of motion that are ordinary differential equations.

For the configurations we will be studying, the action of type IIB supergravity reduces to (in Einstein frame):
\beq
		S_{IIB} = \frac{1}{2\kappa_{10}^2} \left[ \int \sqrt{-g} \left( R - \frac{1}{2} \partial_{\mu} \phi \partial^{\mu} \phi \right) - \frac{1}{2} \int \left( e^{\phi} F_{(3)} \wedge *F_{(3)} \right)\right] \,,
\eeq
while the action for the smeared sources is:
\beq
	S_{\text{sources}} = -T_{D5} \int \Big(e^{\phi/2} \cK_{D5} - C_{(6)}  \Big) \wedge \Xi\,,
	\label{eqn:S_sources}
\eeq
where $\Xi$ is the so-called smearing form that accounts for the distribution of the flavor branes, and $\cK_{D5}$ is the calibration form for the D5-branes. This calibration form can be defined in terms of spinor bilinears (see \cite{Gaillard:2008wt} for the details of this construction), and the fact that it is appearing in \eqref{eqn:S_sources} instead of the usual volume factor is due to supersymmetry. Indeed, a flavor brane being supersymmetric is equivalent to the statement that the pullback of $\cK_{D5}$ on its world-volume equals its volume form.

The introduction of $S_{\text{sources}}$ in \eqref{eqn:S} modifies the equations of motion of type IIB supergravity. In addition, it is responsible for the violation of the Bianchi identity for $F_3$, which allows us to relate the smearing form to the RR flux:
\beq
		\td F_{(3)} = 2 \kappa_{10}^2 T_{D5}\, \Xi\,.
\eeq

In this paper, we decide to look for solutions of \eqref{eqn:S_sources} that are dual to field theories exhibiting a Kutasov-like duality. Kutasov duality is a generalization of Seiberg duality \cite{Seiberg:1994pq}. It relates two four-dimensional $\cN = 1$ gauge theories. One has gauge group $SU(N_c)$, with $N_f$ chiral multiplets in the fundamental representation, and one adjoint chiral superfield $X$ with the following superpotential:
\beq
	W(X) = \Tr \sum_{l=1}^k g_l X^{l+1}\,.
\eeq
where $k$ is an integer. The second gauge theory, related by Kutasov duality to the one we just described, is very similar: it has gauge group $SU(k N_f- N_c)$ with $N_f$ fundamental chiral superfields and one adjoint one $Y$. In addition it has $N_f^2$ mesons. The details of the construction of the mesons and the superpotential for $Y$ in terms of quantities of the first gauge theory can be found in \cite{Kutasov}. It can be generalized to the case where we have multiple generations of adjoint chiral superfields \cite{Abel:2009ty}. We will show the way one can see this Kutasov duality in our supergravity solutions. Especially, we will identify the parameter $k$ of the duality with some gravity quantities.
\medskip

The structure of the paper is as follows: In Section \ref{sec:HxH.ansatz} we find the supergravity differential equations describing branes wrapping Riemann surfaces with higher genus. We will be able to reduce the study of this system of equations to the study of a simple ordinary second-order differential equation. We systematically investigate the solutions of this ODE in Section \ref{sec:HyperSol}. In Section \ref{sec:FT} we critically analyze several features of the dual gauge theory to our brane configuration. We show, among other things, that we do see a realization of Kutasov duality in the supergravity picture. Section \ref{sec:mixed} can be read independently, and can be skipped at first; it deals with a generalization of the ansatz previously used, allowing for more general brane configurations, as well as with the study of its solutions. We did not however study the field theories dual to those additional solutions. Finally we present some conclusions. We further include several appendices filling in some missing details and complementing the main text.

\section{The $\HxH$ ansatz}
\label{sec:HxH.ansatz}

Our goal is to find type IIB supergravity solutions that correspond to D5-branes wrapping Riemann surfaces of higher genus. As we know from the uniformization theorem (see Section \ref{ssec:geometry} for details), these admit a geometric structure modeled on the hyperbolic plane $\mH_2$, this being the reason why we will often refer to these surfaces as hyperbolic two-cycles. As we will argue later (see Section \ref{sec:FT}), our motivation in looking for such configurations is that of finding gravity duals to supersymmetric gauge theories with massless adjoint matter. For the moment, in these first sections, we focus only on the gravity side and the quest for these new type IIB supergravity solutions.

We are interested in finding geometries dual to four-dimensional $\cN=1$ gauge theories. One simple way to achieve this is by endowing the geometries with an $SU(3)$-structure. Additionally it tells us that we should wrap our D5-branes on a two-cycle as mentioned before so that, for energies that appear small compared to the inverse size of the cycle, the six-dimensional theory on the branes reduces to a four-dimensional one. There is a close example that achieves exactly this, which is the so-called Maldacena-N\'u\~nez model \cite{Maldacena:2000yy,Chamseddine:1997nm}. So it is interesting to revisit it as a starting point for motivating the ansatz we will use later (the reader who wants to skip this part can jump directly to \eqref{eqn:H2H3.metric}). In fact, for our present purpose, it is far more appropriate to have a look at a generalization of the MN solution: the one found in \cite{Casero:2006pt} by Casero, N\'u\~nez and Paredes, that has come to be known as the CNP solution, which accounts for the inclusion of dynamical massless flavors into the MN background (see also \cite{Casero:2007jj} for a more precise matching with the dual field theory). Let us recall then how this CNP geometry looks like.

\subsection{The CNP solution}

By wrapping a large number $N_c$ of D5-branes on a two-sphere inside a Calabi-Yau threefold, and adding a smeared set of $N_f$ ($\sim N_c$) D5-branes overlapping with the former along Minkowski space-time, one finds a type IIB supergravity solution dual to an $\cN=1$ $SU(N_c)$ SQCD-like theory with $N_f$ flavors.

In Einstein frame, and with the conventions $\a'=1=g_s$, the metric, RR three-form and dilaton cast as:
\begin{align}
\td s^2 &= e^{2f} \Big[ \td x_{1,3}^2 + e^{2k} \td r^2 + e^{2h} \big( \s_1^2 + \s_2^2 \big) + \frac{e^{2g}}{4} \big( (\w_1 - A_1)^2 + (\w_2 - A_2)^2 \big) + \frac{e^{2k}}{4} (\w_3 - A_3)^2 \Big]\,,\label{eqn:S2S3.metric}\\
F_3&=-\frac{N_c}{4}\bigwedge_i(\w_i - B_i)+\frac{N_c}{4}\sum_i G_i\wedge(\w_i-B_i)-\frac{N_f}{4}\s_1\wedge\s_2\wedge(\w_3-B_3)\,,\\
\f&=4f\,,
\end{align}
where $f,g,h,k$ are all functions of the radial/holographic coordinate $r$; $\s_{1,2}$ parameterize a two-sphere $\mS^2$ and $\w_{1,2,3}$ parameterize a three-sphere $\mS^3$. These $\w_{1,2,3}$ are $\mathfrak{su}(2)$ left-invariant one-forms satisfying the Maurer-Cartan relations:
\beq
\td\w_i=-\frac{1}{2}\eps_{jik}\w_j\wedge \w_k\,.
\label{eqn:su2algebra}
\eeq
The set of $\mS^2$ one-forms $\s_{1,2}$ can be completed with a third one $\s_3$, such that they mimic the $\mS^3$ Maurer-Cartan algebra, $\td\s_i=-\frac{1}{2}\eps_{ijk}\s_j\wedge \s_k$, although they are obviously not independent. The one-forms $A_i,B_i$ entering the fibration and the RR form then read:
\beq
A_{1,2}=a\,\s_{1,2}\,,\quad A_3=\s_3\quad\qquad;\quad\qquad B_{1,2}=b\,\s_{1,2}\,,\quad B_3=\s_3\,,
\eeq
where $a,b$ are also functions of $r$. Finally the two-forms $G_i$ appearing in $F_3$ can be written as a gauge field-strength for $B_i$:
\beq
G_i=\td B_i+\frac{1}{2}\eps_{ijk}B_j\wedge B_k\,.
\eeq
For concreteness, let us show a coordinate representation for the left-invariant one-forms used above. If we choose the usual coordinate system for the $\mS^2$ and $\mS^3$, $\left\{\th_1,\vf_1\right\}$ and $\left\{\th_2,\vf_2,\j\right\}$ respectively, we have:
\beq\begin{aligned}
&\s_1=-\td\th_1\,,\qquad\qquad&&\w_1=\cos\j\,\td\th_2+\sin\j\,\td\vf_2\,,\\
&\s_2=\sin\th_1\,\td\vf_1\,,\qquad\qquad&&\w_2=-\sin\j\,\td\th_2+\cos\j\,\td\vf_2\,,\\
&\s_3=-\cos\th_1\,\td\vf_1\,,\qquad\qquad&&\w_3=\td\j+\cos\th_2\,\td\vf_2\,.
\end{aligned}\eeq
The CNP background is 1/8-supersymmetric and has consequently four Killing spinors that satisfy the following projections:
\beq
\eps=\t_1\,\eps\,,\qquad\Gamma_{12}\eps=\Gamma_{34}\eps\,,\qquad\Gamma_{r345}\eps=\cos\a\,\eps+\sin\a\,\Gamma_{24}\eps\,,
\eeq
where $\t_1$ is the first Pauli matrix, $\a=\a(r)$, and the $\Gamma_{ a_1 a_2\cdots}$ are antisymmetrized products of constant Dirac matrices in the natural vielbein frame for the metric \eqref{eqn:S2S3.metric}:
\beq\bal
e^{x^i} &= e^f \td x^i\,,\quad (i=0,1,2,3)\,,\quad && e^r = e^{f+k} \td r\,, \\
e^1 &= e^{f+h} \s_1\,, && e^2 = e^{f+h} \s_2\,, \\
e^3 &= \frac{e^{f+g}}{2} (\w_1 - A_1)\,, && e^4 = \frac{e^{f+g}}{2} (\w_2 - A_2)\,,\quad & e^5 &= \frac{e^{f+k}}{2} (\w_3 - A_3)\,.\\
\eal\eeq
The functions $f,g,h,k,a,b,\a$ characterizing the background are known\footnote{For general $N_c, N_f$, the full solutions are only known numerically. Only the asymptotic UV and IR behaviors are known analytically.} as the solution of a system of first-order ordinary differential equations, the so-called BPS system. This BPS system can be reduced to a second-order ODE, which we will call ``master equation'' since, once it is solved, all the previous functions follow. This master equation is simpler if we perform the reparameterization of the ansatz that was originally proposed in \cite{HoyosBadajoz:2008fw}. After this reparameterization, the geometry will not be as transparent as in \eqref{eqn:S2S3.metric}, where we can clearly see an $\mS^3$ fibered over an $\mS^2$ (reason why we will refer to this CNP solution as the $\SxS$ case), but in turn, the analytic treatment of the solution is much simpler. The change of variables reads as follows:
\beq\bal
	e^{2h} &= \frac{1}{4} \frac{P^2 - Q^2}{P \cosh \t - Q}\,,\qquad& a &= \frac{P \sinh \t }{P \cosh \t - Q}\,,\qquad&\cos \a &= \frac{P - Q \cosh \t}{P \cosh \t - Q}\,,\\
	e^{2g} &= P \cosh \t - Q\,,&b &= \frac{\s}{N_c}\,,& \sin \a &= - \frac{\sinh \t \sqrt{P^2 - Q^2}}{P \cosh \t - Q}\,,\\
	e^{2k} &= 4 Y\,,& e^{2 \phi} &= \frac{D}{Y^{1/2}(P^2 -Q^2)} \,,\label{eqn:S2S3.change2PQ}
\eal\eeq
where, of course, the new functions $P, Q, Y, \t, \s, D$ depend only on $r$. Note there is one function less than before. This occurs because $\a$ could be written in terms of the others as a consequence of supersymmetry. In these new variables, the CNP solution reads:
\beq\bal
&\s=\tanh\t\,\left(Q+\frac{2N_c-N_f}{2}\right)\,,\\
&\sinh\t\,=\frac{1}{\sinh(2r-2r_0)}\,,\\
&D=e^{2\f_0}\sqrt{P^2-Q^2}\cosh (2r_0)\,\sinh(2r-2r_0)\,,\\
&Y=\frac{1}{8}\left(P'+N_f\right)\,,\\
&Q=\left(Q_0+\frac{2N_c-N_f}{2}\right)\coth(2r-2r_0)+\frac{2N_c-N_f}{2}\left(2r\coth(2r-2r_0)\,-1\right)\,,
\label{eqn:S2S3.sol}
\eal\eeq
where the prime denotes differentiation with respect to $r$, the terms with a zero subindex are constants, and $P$ is the solution of the following second-order differential equation:
\beq
	P'' + (P' + N_f) \left( \frac{P' + Q' + 2 N_f}{P - Q} + \frac{P' - Q' + 2 N_f}{P + Q} - 4 \coth(2r-2r_0)\, \right) = 0\,.
\label{eqn:S2S3.master}
\eeq
We will dub \eqref{eqn:S2S3.master} as the master equation for the $\SxS$ case.

\subsection{The ansatz}

Inspired by \eqref{eqn:S2S3.metric}, we write down an ansatz for a type IIB supergravity solution representing D5-branes wrapping a hyperbolic two-cycle (recall that by this we mean a Riemann surface with genus $g>1$), plus a smeared set of $N_f$ flavor D5-branes. The first guess would be to substitute the $\mS^2$ appearing in \eqref{eqn:S2S3.metric} by an $\mH_2$.\footnote{Recall the Riemann surface can be later obtained from $\mH_2$ by quotienting by a Fuchsian group $\Gamma$, and this leaves locally the same metric as that of $\mH_2$. See appendix \ref{app:quotients}.} However, we know that this $\mS^2$ is not the two-cycle the D5-branes are wrapping. The latter actually involves another $\mS^2$ inside the $\mS^3$ as well \cite{Bertolini:2002yr}. Then it makes sense to think that we need also to substitute the $\mS^3$ by some three-dimensional manifold that can accommodate the hyperbolic two-cycle inside it.

This substitution can be achieved by keeping basically the same ansatz as in the $\SxS$ case:
\begin{align}
\td s^2 &= e^{2f} \Big[ \td x_{1,3}^2 + e^{2k} \td r^2 + e^{2h} \big( \us_1^2 + \us_2^2 \big) + \frac{e^{2g}}{4} \big( (\uw_1 - A_1)^2 + (\uw_2 - A_2)^2 \big) + \frac{e^{2k}}{4} (\uw_3 - A_3)^2 \Big]\,,\label{eqn:H2H3.metric}\\
F_3&=-\frac{\tN_c}{4}\bigwedge_i(\uw_i - B_i)+\frac{\tN_c}{4}\sum_i \underline{G}_i\wedge(\uw_i-B_i)-\frac{\tN_f}{4}\s_1\wedge\s_2\wedge(\uw_3-B_3)\,,\label{eqn:H2H3.F3}\\
\f&=4f\,,\label{eqn:H2H3.dil}
\end{align}
where $f,g,h,k$ are all functions of the radial/holographic coordinate $r$; but now we are using a different set of left-invariant one-forms $\uw_i$, such that they satisfy the following Maurer-Cartan relations:
\beq
\td\uw_1=-\uw_2\wedge \uw_3\,,\quad\td\uw_2=-\uw_3\wedge \uw_1\,,\quad\td\uw_3=+\uw_1\wedge \uw_2\,.
\label{eqn:Halgebra}
\eeq
Notice the flip of the last sign with respect to \eqref{eqn:su2algebra}. This choice will enforce the presence of hyperbolic cycles. We are also using a different set of one-forms $\us_i$, that characterize the $\mH_2$ in the same way as the $\s_i$ characterized the $\mS^2$, and once again mimic the algebra \eqref{eqn:Halgebra} of their $\uw_i$ counterparts: $\td\us_1=-\us_2\wedge \us_3$, $\td\us_2=-\us_3\wedge \us_1$ and $\td\us_3=+\us_1\wedge \us_2$. The one-forms $A_i,B_i$ entering the fibration and the RR form stay as in the $\SxS$ case:
\beq
A_{1,2}=a\,\us_{1,2}\,,\quad A_3=\us_3\quad\qquad;\quad\qquad B_{1,2}=b\,\us_{1,2}\,,\quad B_3=\us_3\,,
\eeq
with $a=a(r)$, $b=b(r)$, but we have to modify slightly the definition of the gauge field-strength:
\beq
\underline{G}_i=\td B_i+\frac{1}{2}\eps_{ijk}B_j\wedge B_k\,,\quad(i=1,2)\quad;\qquad \underline{G}_3=-\left(\td B_3-B_1\wedge B_2\right)\,.
\eeq

\medskip

In what follows, we will use this vielbein base for the metric \eqref{eqn:H2H3.metric}:
\beq\bal
e^{x^i} &= e^f \td x^i\,,\quad (i=0,1,2,3)\,,\quad && e^r = e^{f+k} \td r\,, \\
e^1 &= e^{f+h} \us_1\,, && e^2 = e^{f+h} \us_2\,, \\
e^3 &= \frac{e^{f+g}}{2} (\uw_1 - A_1)\,, && e^4 = \frac{e^{f+g}}{2} (\uw_2 - A_2)\,,\quad & e^5 &= \frac{e^{f+k}}{2} (\uw_3 - A_3)\,.\label{eqn:H2H3.vielbein}
\eal\eeq

Let us exhibit a definite coordinate representation for the one-forms $\uw_i$ and $\us_i$ above. First, if we choose the metric of the Poincar\'e half-plane $\mH_2$ as it is customary: $\td s^2=\frac{\td z_1^2+\td y_1^2}{y_1^2}$, the following one-forms:
\beq
\us_1=-\frac{\td y_1}{y_1}\,,\quad\us_2=-\frac{\td z_1}{y_1}\,,\quad\us_3=-\frac{\td z_1}{y_1}\,,\quad
\label{eqn:us's}
\eeq
play the same role as the one the $\s_i$ played for the $\mS^2$. Note that the $\us_i$ are clearly not independent, as it happened with the $\s_i$.

Then, to specify some coordinate representation of $\uw_i$, we should first know which three-manifold they parameterize. This will be a squashed version of the universal cover of $SL_2(\mR)$, that we will denote by $\SL$, as we discuss in Section \ref{ssec:geometry}. $\SL$ can be built as an $\mS^1$ fiber bundle over $\mH_2$, which shows that a hyperbolic two-cycle can be accommodated inside it. Choosing $z_2,y_2$ for the coordinates of $\mH_2$ as before, and $\j$ as the coordinate for the fiber, the $\uw_i$ read:
\beq
\uw_1=\cos\j\,\frac{\td y_2}{y_2}-\sin\j\,\frac{\td z_2}{y_2}\,,\quad\uw_2=-\sin\j\,\frac{\td y_2}{y_2}-\cos\j\,\frac{\td z_2}{y_2}\,,\quad\uw_3=\td\j+\frac{\td z_2}{y_2}\,.
\label{eqn:uw's}
\eeq
The range of these coordinates $\left\{z_1,y_1,z_2,y_2,\j\right\}$ do not bother us for the moment, since we will eventually take a quotient of both $\mH_2$ and $\SL$ by some freely acting discrete isometry groups $\Gamma$ and $G$ respectively. These quotients need to be taken in order to generate the higher genus surface from $\mH_2$ and a compact space out of $\SL$. They are reflected on the fact that in the ansatz for $F_3$ \eqref{eqn:H2H3.F3}, neither $N_c$ nor $N_f$ appear directly, but rather some related quantities $\tN_c$, $\tN_f$. We will see in Section \ref{ssec:brane.setup} what the relation is.

\subsection{Supersymmetry analysis}
\label{sec:SUSYanalysis}

We want our background \eqref{eqn:H2H3.metric}-\eqref{eqn:H2H3.dil} to possess four supersymmetries. That is, one eighth of the thirty-two supercharges of type IIB supergravity should be preserved. As one can see in \eqref{eqn:H2H3.metric}, our space is of the form $\mathcal{M}_4 \times_w X_6$ where $\mathcal{M}_4$ is four-dimensional Minkowski space, $X_6$ is a six-dimensional manifold and $\times_w$ means a warped product. One way to dictate the preservation of only four supercharges is to impose that our six-dimensional internal manifold $X_6$ be endowed with an $SU(3)$-structure. We are interested in having only the three-form flux $F_{3}$ non-zero, so our $SU(3)$-structure will be parameterized by one two-form $J$ and one three-form $\Omega$. In the basis of \eqref{eqn:H2H3.vielbein}, one can define the $SU(3)$-structure forms as:
\beq
	\bal
		J &= e^r\wedge e^5 + e^1 \wg (\cos \a\, e^2 + \sin \a\, e^4) + e^3 \wg (\sin \a\, e^2 - \cos \a\, e^4)\,, \\
		\W &= (e^r + i e^5) \wg (e^1 + i (\cos \a \, e^2 + \sin \a\, e^4)) \wg (e^3 + i (\sin \a\, e^2 - \cos \a\, e^4))\,,
	\eal
\eeq
where, once again, $\a$ is a function of $r$ only. $G$-structures are a way to express supersymmetry in a geometrical form. So one can write the supersymmetry equations in terms of the $SU(3)$-invariant forms $J$ and $\W$.
The BPS system of first-order differential equations is then given by \cite{Martucci:2005ht}:
\beq
	\bal
		&\td (e^{3f + \phi/2} \W) = 0\,, \qquad\qquad		&&\td (e^{4f} J \wg J) = 0\,, \\
 		&\td (e^{2f - \phi/2}) = 0\,, \qquad\qquad 		&&\td (e^{2f + \phi} J) = - e^{2f + 3\phi/2} *_6 F_{(3)}\,,
 	\eal
\eeq
where $*_6$ indicates the Hodge dual in the internal manifold. In addition, the $SU(3)$-structure also plays a role when writing the action for the flavor branes. Indeed, supersymmetry is equivalent to the $SU(3)$-structure in the case we are studying and the flavor branes are supersymmetric. So it makes sense that the calibration form $\cK_{D5}$ appearing in \eqref{eqn:S_sources} can be written in terms of $SU(3)$-structure forms, namely:
\beq \label{eq:Calibration}
	\cK_{D5} = e^{4f} \td x^0 \wg \td x^1 \wg \td x^2 \wg \td x^3 \wg J\,.
\eeq

The system found from these equations can be obtained from the one found in \cite{Casero:2006pt} by doing the following transformations:
\beq
e^g \rightarrow -i e^g\,,\qquad e^h \rightarrow -i e^h\,,\qquad a\rightarrow -i a\,,\qquad b\rightarrow -i b\,,\qquad N_c\rightarrow\tN_c\,,\qquad N_f\rightarrow\tN_f\,.
\eeq
However, we can directly study it after making the following redefinitions for our functions:
\beq \label{eq:HyperChange}
\bal
	e^{2h} &= -\frac{1}{4} \frac{P^2 - Q^2}{P \cosh \t - Q}\,,\qquad& a &= \frac{P \sinh \t }{P \cosh \t - Q}\,,\qquad&\cos \a &= - \frac{P - Q \cosh \t}{P \cosh \t - Q}\,,\\
	e^{2g} &= -P \cosh \t + Q\,,&b &= \frac{\s}{\tN_c}\,,& \sin \a &= \frac{\sinh \t \sqrt{P^2 - Q^2}}{P \cosh \t - Q}\,,\\
	e^{2k} &= 4 Y\,,& e^{2 \phi} &= \frac{D}{Y^{1/2}(P^2 -Q^2)} \,,\\
\eal\eeq
where of course the new functions $P, Q, Y, \t, \s, D$ depend only on $r$. Note the change of sign in the transformation of $e^{2g}$ and $e^{2h}$ as compared to \eqref{eqn:S2S3.change2PQ}.

In terms of those new functions, the BPS system can be written as
\beq
	{\allowdisplaybreaks\bal
		&P' = 8Y - \tN_f \,,\\
		&\left( \frac{Q}{\cosh \t} \right)' = \frac{2\tN_c - \tN_f}{\cosh^2 \t}\,, \\
		&\frac{\td}{\td r} \log \left( \frac{D}{\sqrt{P^2 -Q^2}} \right) = 2 \cosh \t\,, \\
		&\frac{\td}{\td r} \log \left( \frac{D}{\sqrt{Y}} \right) = \frac{16 Y P}{P^2 - Q^2}\,, \\
		&\t' + 2 \sinh \t = 0 \,,\\
		&\s = \tanh \t \left( Q + \frac{2\tN_c - \tN_f}{2} \right)\,.
	\eal}%
\eeq
This BPS system is identical (barring the tildes in $\tN_c,\tN_f$) to the one of the $\SxS$ case, and it is solved in the same manner:
\beq\bal
&\s=\tanh\t\,\left(Q+\frac{2\tN_c-\tN_f}{2}\right)\,,\\
&\sinh\t\,=\frac{1}{\sinh(2r-2r_0)}\,,\\
&D=e^{2\f_0}\sqrt{P^2-Q^2}\cosh (2r_0)\,\sinh(2r-2r_0)\,,\\
&Y=\frac{1}{8}\left(P'+\tN_f\right)\,,\\
&Q=\left(Q_0+\frac{2\tN_c-\tN_f}{2}\right)\coth(2r-2r_0)+\frac{2\tN_c-\tN_f}{2}\left(2r\coth(2r-2r_0)\,-1\right)\,.
\eal\label{eqn:whatever}\eeq
And we then remain with a second-order differential equation:
\beq
	P'' + (P' + \tN_f) \left( \frac{P' + Q' + 2 \tN_f}{P - Q} + \frac{P' - Q' + 2 \tN_f}{P + Q} - 4 \coth(2r-2r_0)\,\right) = 0\,.
\label{eqn:H2H3.master}
\eeq
The search for solutions boils down to solving this master equation\footnote{It can be checked that, as expected, the fulfillment of the equations of motion of type IIB supergravity is implied by the fulfillment of this master equation and the Bianchi identity violation \cite{Koerber:2007hd}.}, which is, apart from the change $N_f\to\tN_f$, identical to the master equation of the $\SxS$ case \eqref{eqn:S2S3.master}. However, it is important to notice that in the case at hand, in order for the transformation \eqref{eq:HyperChange} and the solution \eqref{eqn:whatever} to be well-defined, we are looking for solutions such that
\beq \label{eq:conditionsHxH}
		Q \geq P \cosh \t\,, \;\;\;\;\; P^2 \geq Q^2\,, \;\;\;\;\; P' + \tN_f \geq 0\,,
\eeq
which makes the solutions of this $\HxH$ case behave very differently from their $\SxS$ relatives.

\subsection{Brane setup}
\label{ssec:brane.setup}

Let us briefly discuss the brane configuration our background \eqref{eqn:H2H3.metric}-\eqref{eqn:H2H3.F3} is describing. The idea is that we have $N_c$ D5-branes (the so-called color branes), wrapping a hyperbolic two-cycle inside a Calabi-Yau threefold. When we take this number $N_c$ to be very large, plus a near-horizon limit, the Calabi-Yau threefold undergoes a geometric transition and the branes dissolve into flux \cite{Vafa:2000wi}. The resulting internal manifold preserves the $SU(3)$-structure, and topologically is an interval times $\dfrac{\mH_2}{\Gamma}\times\dfrac{\SL}{G}$, as sketched below:
\begin{center}\begin{tabular}{|c|c|c|}
\hline
$\left[r_{IR},r_{UV}\right]$ & $\mH_2/\Gamma$ & $\SL/G$\\
\hline
$r$ & $z_1,y_1$ & $z_2,y_2,\j$\\
\hline
\end{tabular}\end{center}
From the general geometric transition picture, one would expect to find a vanishing hyperbolic two-cycle in the IR, which by analogy with what happens in the MN solution should read\footnote{Actually there are two equivalent two-cycles, the other one being defined by $z_1=-z_2$, $y_1=y_2$, $\j=\p$. It can be checked that these two-cycles are indeed vanishing in the IR when we remove the flavors from the solution. See Section \ref{ssec:Nf=0}.} $z_1=z_2$, $y_1=-y_2$, $\j=\p$; and a blown-up three-cycle pervaded by the three-from flux. A good choice for this three-cycle is $\SL$, and what remains from the initial $N_c$ branes is the flux quantization condition:
\beq
-N_c=\frac{1}{2\k^2_{(10)}T_{D5}}\int_{\SL}\imath^{*}(F_{(3)})=-\frac{\tN_c\,\vol{\SL}}{2\p^2}\,,
\eeq
where we are abusing notation and denoting by $\SL$ the actual appropriate compact quotient $\SL/G$. The volume is to be understood as taking into account possible winding effects. The inclusion of this submanifold in the ten-dimensional background, used for the pullback, has been denoted by $\imath$. Note that from here we get:
\beq
\tN_c=\frac{2\p^2}{\vol{\SL}}N_c.
\label{eqn:tNc}
\eeq

As for the relation between $\tN_f$ and $N_f$, it can be obtained by looking at the violation of the Bianchi identity. As in the CNP solution, the $\tN_f$ in \eqref{eqn:H2H3.F3} is accounting for a set of $N_f$ D5-branes extended along $(r,\j)$ plus Minkowski coordinates\footnote{It is easy to see that this six-cycle is $\k$-symmetric, for instance by looking at the calibration six-form \eqref{eq:Calibration}, and checking that $\pbi{\cK_{D5}}=\o_{\vol{\pbi{g}}}$.} (with the transverse coordinates being constant), and homogeneously smeared over the space transverse to them. Thus, the violation of the Bianchi identity should read:
\beq
\td F_{3}=-2\k^2_{(10)}T_{D5}\frac{N_f}{\vol{\mH_2\times\mH_2}}\o_{\vol{\mH_2\times\mH_2}}\,,
\label{eqn:dF3}
\eeq
where by $\o_{\textrm{Vol}}$ we denote the volume form,  and we are abusing notation once again by having $\mH_2$ stand for the quotient $\mH_2/\Gamma$. There are two $\mH_2$'s in \eqref{eqn:dF3}. Recalling the sketchy table above, one is characterized by $(z_1,y_1)$, and the other one, being the base space of $\SL$ when thought as a line bundle over $\mH_2$, is characterized by the $(z_2,y_2)$ coordinates. As we will see later, it is possible to take simultaneously the same quotient $\mH_2/\Gamma$ in both of them.

From \eqref{eqn:H2H3.F3} we obtain:
\beq
\td F_3=-\frac{\tN_f}{4}\o_{\vol{\mH_2\times\mH_2}}\,,
\eeq
and the comparison with the previous equation \eqref{eqn:dF3} yields the relation we were looking for:
\beq
\tN_f=\frac{(4\p)^2}{\vol{\mH_2}^2}N_f.
\label{eqn:tNf}
\eeq

\subsection{A geometrical remark}
\label{ssec:geometry}

The way we substituted the $\mS^2$ wrapped by the D5-branes in the CNP solution (recall this $\mS^2$ was extended along both the topological two-sphere and three-sphere present in this solution) by a Riemann surface of genus $g>1$, ${\cal C}_g$, was by replacing in \eqref{eqn:S2S3.metric} the metrics of the two-sphere and three-sphere by their ``hyperbolic analogues'':
\beq\begin{aligned}
\td s_{\mS^2}^2=\s_1^2+\s_2^2\,&\to\,\td s_{\mH_2}^2=\us_1^2+\us_2^2\,,\\
\td s_{\mS^3}^2=\w_1^2+\w_2^2+\w_3^2\,&\to\,\td s_{\SL}^2=\uw_1^2+\uw_2^2+\uw_3^2\,,
\end{aligned}\label{eqn:sph2hyp}\eeq
where the one-forms $\s_i,\,\us_i,\,\w_i,\,\uw_i$ have been defined in the previous subsections. One can notice that the metrics on the right-hand side of \eqref{eqn:sph2hyp} represent non-compact spaces. The way to get a hyperbolic compact space out of them is by performing a quotient by a discrete subgroup of isometries. Such a quotient will leave locally the very same metrics of \eqref{eqn:sph2hyp}, which will be therefore the metrics we have to use for ${\cal C}_g$ and for the $\mS^1$ fiber bundle over ${\cal C}_g$ respectively. How to perform this quotient is not important for the supergravity analysis, and only some details of it are needed for the matching with the field theory, which have been moved to appendix \ref{app:quotients} not to deviate the reader. This construction of subspaces as quotients by isometries of a bigger space is well-known in Geometry, and from it we can deduce  that in our case these bigger spaces are $\mH_2$ and $\SL$ respectively. For the sake of completeness, we comment a few words on this topic.

All closed (compact and with an empty boundary) smooth two-manifolds can be given a metric of constant curvature. The uniformization theorem for surfaces provides a way to realize this construction in terms of a so-called geometric structure. A geometric structure on a manifold $M$ is a diffeomorphism between $M$ and a quotient space $X/\Gamma$, where $X$ is what one calls a model geometry, and $\Gamma$ is a group of isometries, such that the projection $X\mapsto X/\Gamma$ is a covering map. In the case of two-manifolds, there are three model geometries (homogeneous and simply connected spaces with a ``nice" metric): the two-sphere $\mS^2$, the Euclidean space $E^2$, and the hyperbolic plane $\mH_2$. Any surface with genus $g>1$ is obtained from the latter (see for instance \cite{Peter}).

It is natural to ask whether there exists a similar classification in three dimensions. This question has only been recently, and positively!, answered by G. Perelman\footnote{Perelman's works have become famous because of proving the Poincar\'e conjecture, which says that the only simply connected three-manifold that exists is the three-sphere $\mS^3$, up to diffeomorphisms; this result however was just a corollary of the much stronger statement he proved, the Thurston geometrization conjecture, which classifies all the possible geometric structures on three-manifolds.}, who has proved the Thurston geometrization conjecture \cite{Perelman}. One could naively think that the model geometries in three dimensions are in correspondence with the two-dimensional ones: $\mS^3$, $E^3$ and $\mH_3$. But it is easy to see that these three are not enough since all of them are isotropic, and there are three-manifolds like $\mS^2\times\mathbb{R}$ that are not. In 1982 W. Thurston proposed eight model geometries for the classification of three-manifolds, and proved that a large part of them admitted a geometric structure modeled on these eight geometries. The classification in three dimensions is more complicated than in two dimensions since not all three-manifolds admit a geometric structure, but it is always possible to ``cut any three-manifold into pieces'' such that each of them does admit a geometric structure. This is the content of the geometrization conjecture. We found that a good account of these topics can be read in \cite{Peter}; despite not being completely up-to-date, it deals with a lot of the mathematical constructions we are using.

It is clear that the construction of a geometric structure is appealing to us, since the manifold parameterized by the $\uw_i$'s in \eqref{eqn:H2H3.metric} will be precisely realized as a quotient of a model geometry by a discrete group of isometries. In order to know which of the eight model geometries we are dealing with, we can resort to the relation between these eight geometries and the Bianchi groups: seven of the eight geometries can be realized as a simply-connected three-dimensional Lie group (which were classified by Bianchi) with a left-invariant metric. From this construction (see for instance \cite{Bergshoeff:2003ri} for details) it follows that the metric
\beq
\td s^2=\left(\uw_1\right)^2+\left(\uw_2\right)^2+\left(\uw_3\right)^2\,,
\label{eqn:metric.SL2}
\eeq
corresponds to the Thurston model geometry $\SL$, since the algebra of the $\uw_i$'s relates to the type VIII Bianchi algebra.

\section{Solutions for the case $\mH_2 \times \SL$}
\label{sec:HyperSol}

We have not been able to find a general analytical solution of the master equation \eqref{eqn:H2H3.master}. Of course it is easy to find numerical solutions, but no matter what values we use for the initial conditions, the solutions always seem to exist only on a finite interval $\left[r_0,r_{UV}\right]$. This issue cannot be resolved by a redefinition of the radial coordinate, since the invariant length $\int_{r_0}^{r_{UV}} \td r \sqrt{g_{rr}}$ will be finite for all the solutions. We will identify $r_0$ with the deep IR, and $r\to r_{UV}$ with the UV. This identification will be made precise in Section \ref{ssec:Landau}.\footnote{Notice that $r_{UV}$ denotes the place in the geometry where our solutions stop being valid. It is the furthest point along the RG flow we can probe in the dual field theory.}

Despite the fact that we only found full solutions numerically, we were able to get analytic expansions both in the IR and in the UV . Actually, as we will see, this will be enough to extract all the physically relevant information (about the dual field theory) we want.

Below we present different expansions that correspond to different solutions of the case $\mH_2 \times \SL$. As we prove in appendix \ref{sec:badUV}, because of the constraints \eqref{eq:conditionsHxH}, it is not possible to obtain solutions for this case that extend all the way to infinity. We are restricted to having the end of the space at a finite position $r_{UV}$ in the radial coordinate. Following the arguments made in \cite{HoyosBadajoz:2008fw} for the possible types of IR and UV expansions,  we found one expansion for the IR situated at $r = r_0 > - \infty$ and three different expansions for the UV situated at $r = r_{UV} < \infty$. Restricting ourselves to Frobenius series, it seems that no other consistent expansions can be found. Without loss of generality, we choose $r_{UV} = 0$, so we automatically have $r_0 < 0$.

In addition to presenting each time the solution for the function $P$, we are also going to translate the results back to the original functions $a$, $g$, $h$, $k$ and $\phi$ in order to make it easier to get an idea of the background and to compare with other results in the literature.

\subsection{Expansions in the IR}
\label{ssec:IRexp}

Let us first start by describing the unique infrared expansion, around $r = r_0$. For $Q$ not to have a pole there\footnote{Notice that this condition follows from the constraint on $P$ and $Q$ for this case: if $Q$ has a pole, $P$ must have a pole too, with negative residue, but this is not possible to achieve for finite $r$ because of the $P' \geq -\tN_f$ constraint.}, one needs to impose first $Q_0 = -\frac{2\tN_c - \tN_f}{2} ( 1+2r_0)$. Then one finds that the expansion for the function $P$ is:
\beq
	\bal
		P = &P_0 - \tN_f (r-r_0) + \frac{4}{3} c_+^3 P_0^2 (r-r_0)^3 - 2 c_+^3 \tN_f P_0 (r-r_0)^4 + \frac{4}{5} c_+^3 \left(\tN_f^2 + \frac{4}{3} P_0^2 \right) (r-r_0)^5 \\
		&+ \cO \left( (r-r_0)^6 \right)\,,
	\eal
	\label{eqn:P.IR}
\eeq
where $P_0$ and $c_+$ are free constants that need to obey $P_0 < 0$ and $c_+ > 0$, in order to satisfy the consistency conditions \eqref{eq:conditionsHxH} imposed on the solutions of the master equation.
The functions in the metric are then
\beq
	\bal
		e^{2h} &= -\frac{P_0}{2} (r-r_0) + \frac{1}{2} \tN_f (r-r_0)^2 + \frac{2}{3} P_0 (r-r_0)^3 + \cO \left( (r-r_0)^4 \right)\,, \\
		e^{2g} &= -\frac{P_0}{2} (r-r_0)^{-1} + \frac{\tN_f}{2} - \frac{2}{3} P_0 (r-r_0) + \cO \left( (r-r_0)^2 \right)\,, \\
		e^{2k} &= 2 c_+^3 P_0^2 (r-r_0)^2 - 4 c_+^3 \tN_f P_0 (r-r_0)^3 + \frac{2}{3} c_+^3 (3 \tN_f^2 +4 P_0^2) (r-r_0)^4 + \cO \left( (r-r_0)^5 \right)\,, \\
		e^{4\f - 4\f_{IR}} &= 1 + \frac{4\tN_f}{P_0} (r-r_0) +\frac{10 \tN_f^2}{P_0^2} (r-r_0)^2 + \left( \frac{20 \tN_f^3}{P_0^3} - \frac{8\tN_f}{3P_0}-\frac{8c_+^3 P_0}{3} \right) (r-r_0)^3 \\
		&\qquad + \cO \left( (r-r_0)^4 \right)\,, \\
		a &= 1 - 2 (r-r_0)^2 - \frac{4}{3P_0} (\tN_f -2 \tN_c) (r-r_0)^3 + \cO \left( (r-r_0)^4 \right)\,. \\
	\eal
	\label{eqn:HxH.IR}
\eeq
Looking at these expressions, one notices that in the IR (at $r=r_0$) the dilaton is finite, $e^{2h}$ and $e^{2k}$ go to $0$, while $e^{2g}$ goes to infinity. The issue of the singularity of the solutions in the IR will be addressed later in Section \ref{ssec:Nf=0}.
Let us now present the different possibilities for the UV.

\subsection{Expansions in the UV}
\label{ssec:UVexpansions}

In this section, we present three different possibilities for the UV expansions, that we can group into two classes, class I and class II, for reasons that become apparent when we look at the behavior of the metric functions in each of them. The interpretation of the different UV's is discussed in Section \ref{ssec:RG}. As we previously mentioned, all the UV's happen at finite $r_{UV}$, that we can choose to be $r_{UV}=0$. So in the following, the expansions are around $0$ and for $r < 0$. As we are looking for a solution that has a space ending in $r=r_{UV}$, we search solutions where some function in the metric either goes to zero, or to infinity at $r_{UV}$. Each of the following expansions will have a different function having this behavior.

\begin{figure}
\centering
\begin{tabular}{l r}
\includegraphics[width=0.45\textwidth]{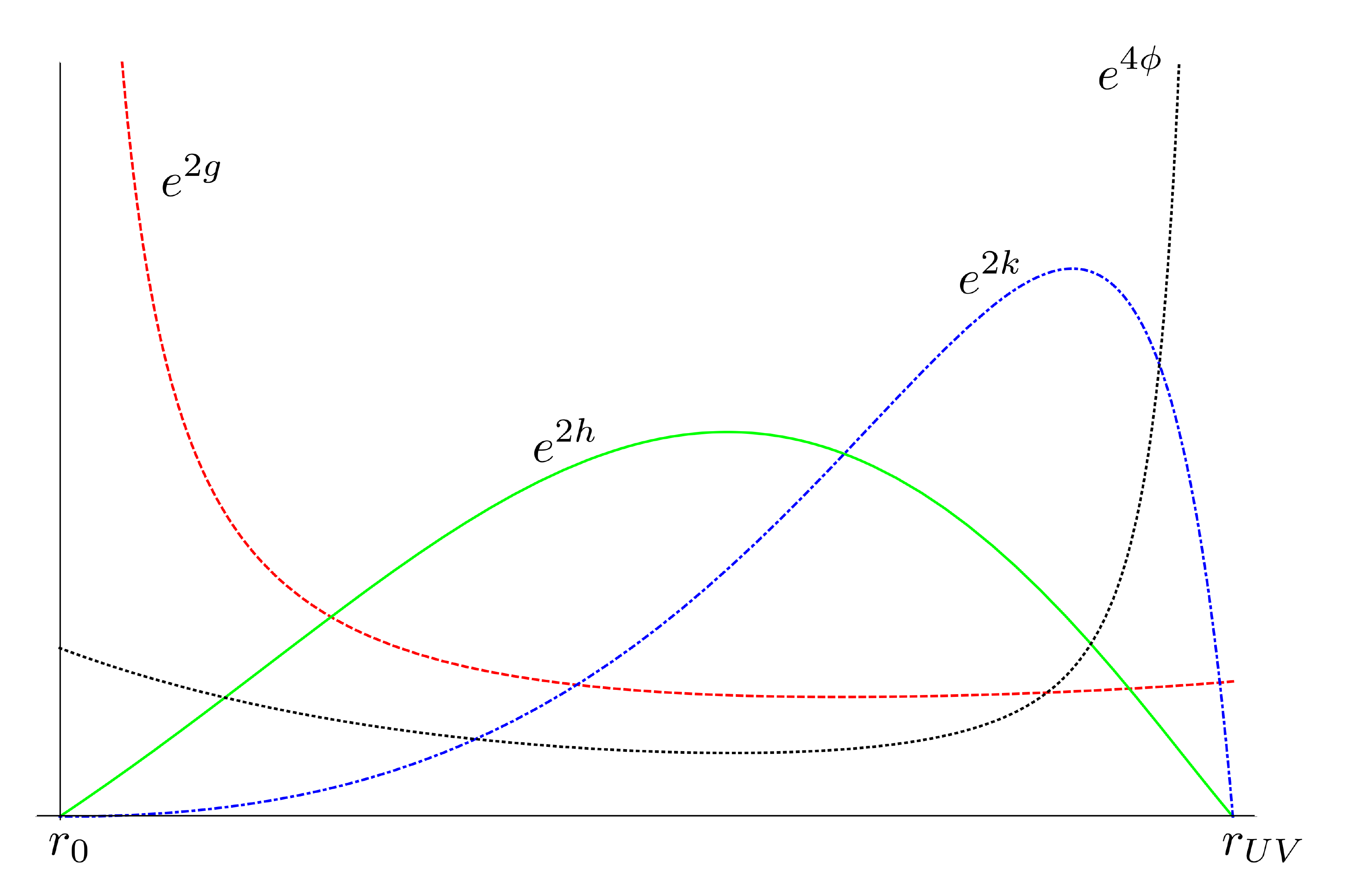} & \includegraphics[width=0.45\textwidth]{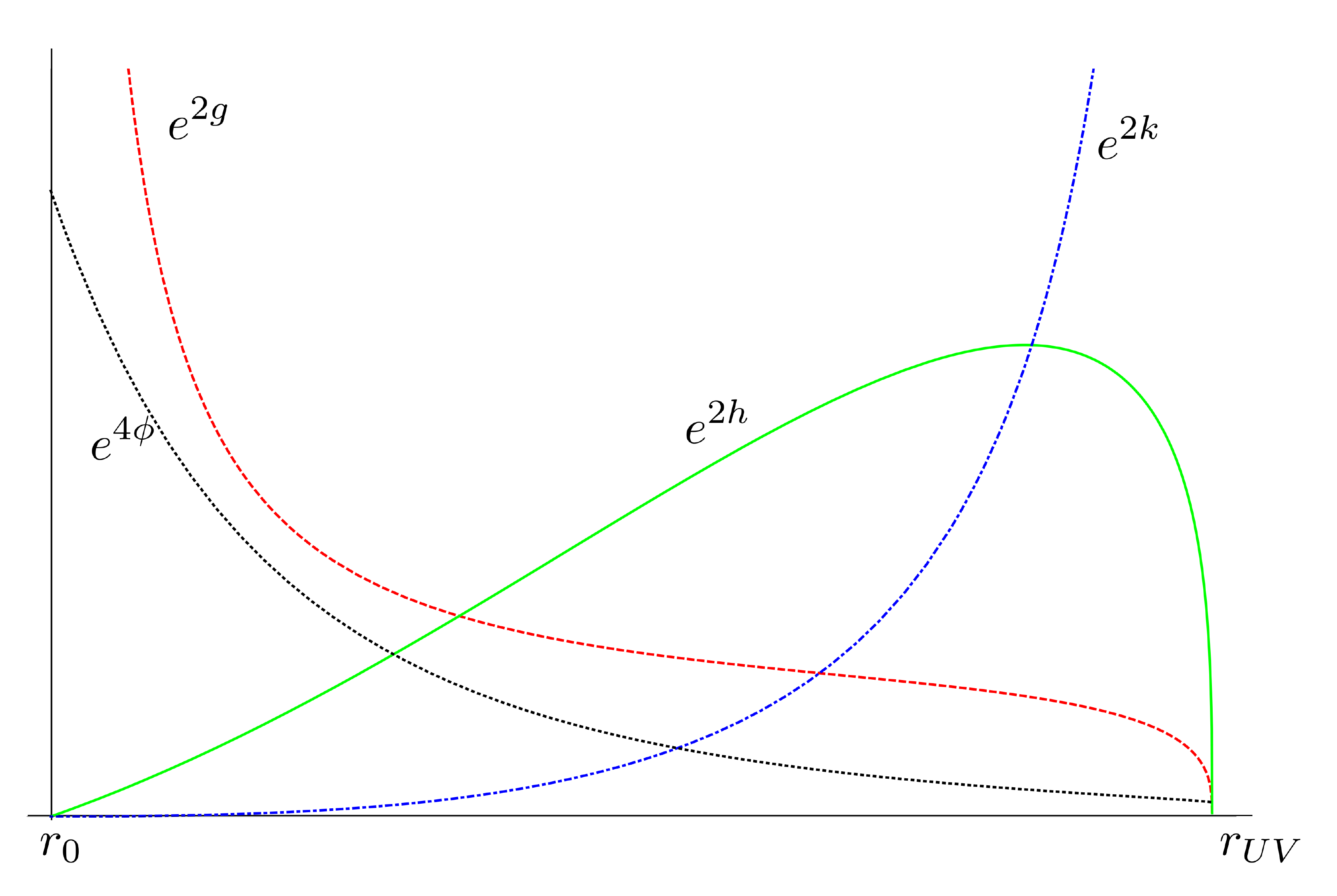}
\end{tabular}
\caption{Plots of the functions $e^{2g}$, $e^{2h}$, $e^{2k}$ and $e^{4\phi}$. On the left, the plots are of class I solutions, while on the right they are of class II.}
\label{fig:plotsUV}
\end{figure}

Let us note that one can find numerical solutions interpolating between the previous IR and each of the following UV's (see Figure \ref{fig:plotsUV}), so we are still working with $Q_0 = -\frac{2\tN_c - \tN_f}{2} ( 1+2r_0)$. Then we can expand $Q$ as:
\beq
	Q = b_0 + b_1 r + b_2 r^2 + \cO \left( r^3 \right)\,,
\eeq
where
\beq
	\bal
		b_0 &= \frac{1}{2} (2 \tN_c -\tN_f) \big( 2 r_0 \coth(2r_0) - 1 \big)\,, \\
		b_1 &= \frac{1}{2} (2\tN_c -\tN_f) \frac{4 r_0 - \sinh(4r_0)}{\sinh^2(2r_0)}\,, \\
		b_2 &= (4 \tN_c - 2 \tN_f) \frac{2 r_0 \cosh(2 r_0) - \sinh(2 r_0)}{\sinh^3(2r_0)}\,.
	\eal
\eeq
Let us now detail the three different expansions and give their domain of validity.

\paragraph{First UV}

The first possible expansion for $P$ is:
\beq
	\bal
		P = &\,Q + h_1 (-r)^{1/2} + \frac{1}{6 b_0} \big( -h_1^2 + 12 b_0 (b_1 + \tN_f) \big) (-r) \\
		&+ \frac{h_1}{72 b_0^2} \big(5 h_1^2 - 6 b_0 ( 5b_1 + 2\tN_f) + 72 b_0^2 \coth(2r_0) \big) (-r)^{3/2} + \cO \left( (-r)^2 \right)\,.
	\eal
\eeq
With this, the functions in the metric are
\begin{eqnarray}
		&&e^{2h} = \frac{h_1}{2 + \coth(r_0) + \tanh(r_0)} (-r)^{1/2} \nn\\
		&&\;\;\;\;\qquad + \frac{h_1^2 + 6 b_0 (b_1+\tN_f) + \coth(2r_0) \big(-2h_1^2 + 6b_0 ( b_1+\tN_f) \big)}{6 b_0 \big(1+\coth(2r_0) \big)^2} (-r) + \cO \left( (-r)^{3/2} \right)\,, \nn\\
		&&e^{2g} = b_0 \big(1+ \coth(2r_0) \big) + h_1 \coth(2r_0) (-r)^{1/2}  + \cO \left( (-r) \right)\,, \nn\\
		&&e^{2k} = -\frac{h_1}{4} (-r)^{-1/2} + \frac{h_1^2 - 6b_0 (b_1+ \tN_f)}{12 b_0} + \cO \left( (-r)^{1/2} \right)\,, \nn\\
		&&e^{4\f-4\f_{UV}} = 1 - \frac{4 (b_1 + \tN_f)}{h_1} (-r)^{1/2} + \cO \left( (-r) \right)\,, \nn\\
		&&a = \cosh(2r_0) - \sinh(2r_0) + \frac{h_1}{b_0 \big( \sinh(2r_0) + \cosh(2r_0) \big)^2} (-r)^{1/2} + \cO \left( (-r) \right)\,.
	\label{eqn:1stUV}
\end{eqnarray}
This is only valid for $\tN_f > 2 \tN_c$ (which gives $b_0 < 0$) and $h_1 < 0$. We have $b_1 + \tN_f > 0$ so the dilaton decreases towards the UV and is finite. We also have
\beq
	e^{4\f_{IR} - 4\f_{UV}} = -\frac{b_0 h_1^2}{c_+^3 P_0^4 \sinh^2(2r_0)}\,.
\eeq
We have $e^{2h}$ going to $0$ while $e^{2k}$ goes to infinity at the UV.

\paragraph{Second UV}

We present now the second possibility for the UV. The expansion for $P$ in that case is
\beq
	\bal
		P = &-Q + h_1 (-r)^{1/2} + \frac{1}{6 b_0} \big( h_1^2 + 12 b_0 (\tN_f - b_1) \big) (-r) \\
		&+ \frac{h_1}{72 b_0^2} \big(5 h_1^2 - 6 b_0 ( 5b_1 - 2\tN_f) + 72 b_0^2 \coth(2r_0) \big) (-r)^{3/2} + \cO \left( (-r)^2 \right)\,.
	\eal
\eeq
Looking at the metric, it gives that
\beq
	\bal
		e^{2h} &= \frac{h_1}{- 2 + \coth(r_0) + \tanh(r_0)} (-r)^{1/2} \\
		&\qquad + \frac{h_1^2 + 6 b_0 (b_1 - \tN_f) + 2 \coth(2r_0) \big(h_1^2 + 3 b_0 (\tN_f - b_1) \big)}{6 b_0 \big(-1+\coth(2r_0) \big)^2} (-r) + \cO \left( (-r)^{3/2} \right)\,, \\
		e^{2g} &= b_0 \big(1 - \coth(2r_0) \big) + h_1 \coth(2r_0) (-r)^{1/2}  + \cO \left( (-r) \right)\,, \\
		e^{2k} &= -\frac{h_1}{4} (-r)^{-1/2} + \frac{-h_1^2 + 6b_0 (b_1 - \tN_f)}{12 b_0} + \cO \left( (-r)^{1/2} \right)\,, \\
		e^{4\f-4\f_{UV}} &= 1 + \frac{4 (b_1 - \tN_f)}{h_1} (-r)^{1/2} + \cO \left( (-r) \right)\,, \\
		a &= \cosh(2r_0) + \sinh(2r_0) + \frac{h_1}{b_0 \big(- \sinh(2r_0) + \cosh(2r_0) \big)^2} (-r)^{1/2} + \cO \left( (-r) \right)\,.
	\eal
	\label{eqn:2ndUV}
\eeq
This is only valid for $\tN_f < 2 \tN_c$ (which gives $b_0 > 0$) and $h_1 < 0$. In that case we have $-\tN_f < b_1 - \tN_f < 2 \tN_c -2 \tN_f$. So if $\tN_c < \tN_f$, $b_1 - \tN_f < 0$ and the dilaton decreases towards the UV. Otherwise, if $\tN_f < \tN_c$, $b_1 - \tN_f$ can be positive or negative depending on the value of $r_0$. So the dilaton either decreases or increases towards the UV. In any case, we have
\beq
	e^{4\f_{IR} - 4\f_{UV}} = \frac{b_0 h_1^2}{c_+^3 P_0^4 \sinh^2(2r_0)}\,.
\eeq
We also have $e^{2h}$ going to $0$ while $e^{2k}$ goes to infinity at the UV. We see that the qualitative behavior of the metric functions in this UV is the same as that in the first UV. It makes sense then to group them under one common class, that we call class I.

\paragraph{Third UV}

We now write the last possibility for the UV. The expansion for $P$ is
\beq
	\bal
		P = &-b_0 + \tN_f (-r) + P_2 (-r)^2 \\
		&+ \frac{P_2}{3 b_0 (\tN_f - b_1)} \big( b_1^2 - \tN_f^2 + 2 b_0 ( b_2 - 3 P_2) - 8 b_0 (b_1 - \tN_f) \coth(2r_0) \big) (-r)^3  + \cO \left( (-r)^4 \right)\,.
	\eal
\eeq
This leads to
\beq
	\bal
		e^{2h} &= \frac{b_1 - \tN_f}{2 - \coth(r_0) -\tanh(r_0)} (-r) + \cO \left( (-r)^2 \right)\,, \\
		e^{2g} &= b_0 \big( 1 - \coth(2r_0) \big) + \big(2 b_0 \coth^2(2r_0) - 2 b_0 + \tN_f \coth(2r_0) - b_1) (-r) + \cO \left( (-r)^2 \right)\,, \\
		e^{2k} &= - P_2 (-r) \\
		&\quad - \frac{P_2}{2 b_0 (\tN_f - b_1)} \big( b_1^2 - \tN_f^2 + 2 b_0 ( b_2 - 3 P_2) - 8 b_0 (b_1 - \tN_f) \coth(2r_0) \big) (-r)^2 + \cO \left( (-r)^3 \right)\,, \\
		e^{4\f} &= \frac{c_+^3 P_0^4}{8} e^{4\f_0} \frac{8 \cosh^2(r_0) \sinh^2(r_0)}{b_0 P_2 (\tN_f - b_1)} (-r)^{-2} \\
		&\quad + \frac{c_+^3 P_0^4}{8} e^{4\f_0} \frac{2\big(\tN_f^2 - b_1^2 + 2 b_0 (P_2 -b_2)\big) \sinh^2(2r_0)}{b_0^2 P_2 (b_1 - \tN_f)^2} (-r)^{-1} + \cO \left( (-r)^0 \right)\,, \\
		a &= \cosh(2r_0) + \sinh(2r_0) + \frac{\tN_f - b_1 + 2 b_0 \big( -1 + \coth(2r_0) \big)}{b_0 \sinh(2r_0) \big(-1 + \coth(2r_0) \big)^2} (-r) + \cO \left( (-r)^2 \right)\,.
	\eal
	\label{eqn:3rdUV}
\eeq
This case is valid only for $\tN_f < 2 \tN_c$ (which gives $b_0 > 0$), $P_2 < 0$ and $b_1 - \tN_f > 0$. This second condition requires $\tN_f < \tN_c$ and depends on the value of $r_0$ (see previous section). For this UV, $e^{2h}$ and $e^{2k}$ both go to $0$ while the dilaton diverges. Notice that this is a qualitatively very different UV behavior than the one we found in the UV's of class I. That is why we put the third UV in a different class: class II. Figure \ref{fig:plotsUV} shows the difference of behavior of the functions in the metric between the two classes of solutions.

\subsection{Comments on the IR singularity}
\label{ssec:Nf=0}

In order to know whether the solutions for the $\HxH$ case are singular or not in the IR, we can look at the behavior of several curvature invariants around $r =r_0$:
\beq
	\bal
		R \,=\, &\frac{\tN_f^2 e^{-\phi_0/2}}{2c_+^3 P_0^4} (r-r_0)^{-2} + \frac{7\tN_f^3 e^{-\phi_0/2}}{4c_+^3 P_0^5} (r-r_0)^{-1} \\
		&+ e^{-\phi_0/2} \frac{189 \tN_f^4 -16 \tN_f (8\tN_c P_0^2 - 9 c_+^3 P_0^4) + 128 \tN_c^2 P_0^2}{48 c_+^3 P_0^6} + \cO \left( (r-r_0) \right)\,,\\
		R_{\m\n\r\s}R^{\m\n\r\s}\,=\,&\frac{20 e^{-\phi_0}}{c_+^6 P_0^4} (r-r_0)^{-8} + \frac{52\tN_f e^{-\phi_0}}{c_+^6 P_0^5} (r-r_0)^{-7} +\cO \left( (r-r_0)^{-6} \right)\,.
	\eal
\eeq
From that, one can see that a generic solution is indeed singular in the IR, since the Ricci scalar $R \sim (r-r_0)^{-2}$. This was to be expected since we deal here with backreacting massless flavors. Indeed, in our setup, we smear D5-branes that are extended in the radial coordinate $r$ from $r=r_0$ to $r=r_{UV}$. As the branes extend all the way to the IR, at $r=r_0$, their density diverges. Thus they must create a curvature singularity in the space. Notice though that this is a good singularity in the sense that the metric component $g_{tt} = e^{\phi/2}$ is bounded \cite{Maldacena:2000mw}, but since $P_0 <0$, $g_{tt}$ grows towards the IR. We will comment on this point in Section \ref{ssec:WLoops}. 

However, in the unflavored case $\tN_f = 0$, we see that the Ricci scalar goes to a constant in the IR, meaning that the solution is better behaved than the flavored one. The same happens for $R_{\m\n}R^{\m\n}$. Indeed, the problem of the infinite density of branes is not present anymore since we do not consider the addition of sources. Nevertheless, the solution is still singular, as one can see by looking at $R_{\m\n\r\s}R^{\m\n\r\s}$. This singularity could have been expected because of the presence of vanishing higher genus manifolds (which contain non-contractible cycles) in the deep IR. It is a ``better'' singularity than the flavored one since this time, $g_{tt}$ decreases towards the IR. The field theories dual to both the flavored and the unflavored cases are going to be studied in the following section.

\section{Field Theory}
\label{sec:FT}

In this section we would like to interpret several features of our $\HxH$ solution
in the gauge/gravity correspondence picture. We argue that the field theory dual is
of the SQCD-type plus adjoint matter charged under the gauge field and
self-interacting through a dangerously irrelevant polynomic superpotential, and
correspondingly displays a Kutasov-like duality. Notice that this interpretation is only valid for energies smaller than the inverse size of the cycle wrapped by the branes. Moreover, we compute several
observables of the field theory, that give us some insight on its IR and UV
behaviors.

\subsection{RG flow}
\label{ssec:RG}

\begin{figure}
\centering
\includegraphics[width=0.6\textwidth]{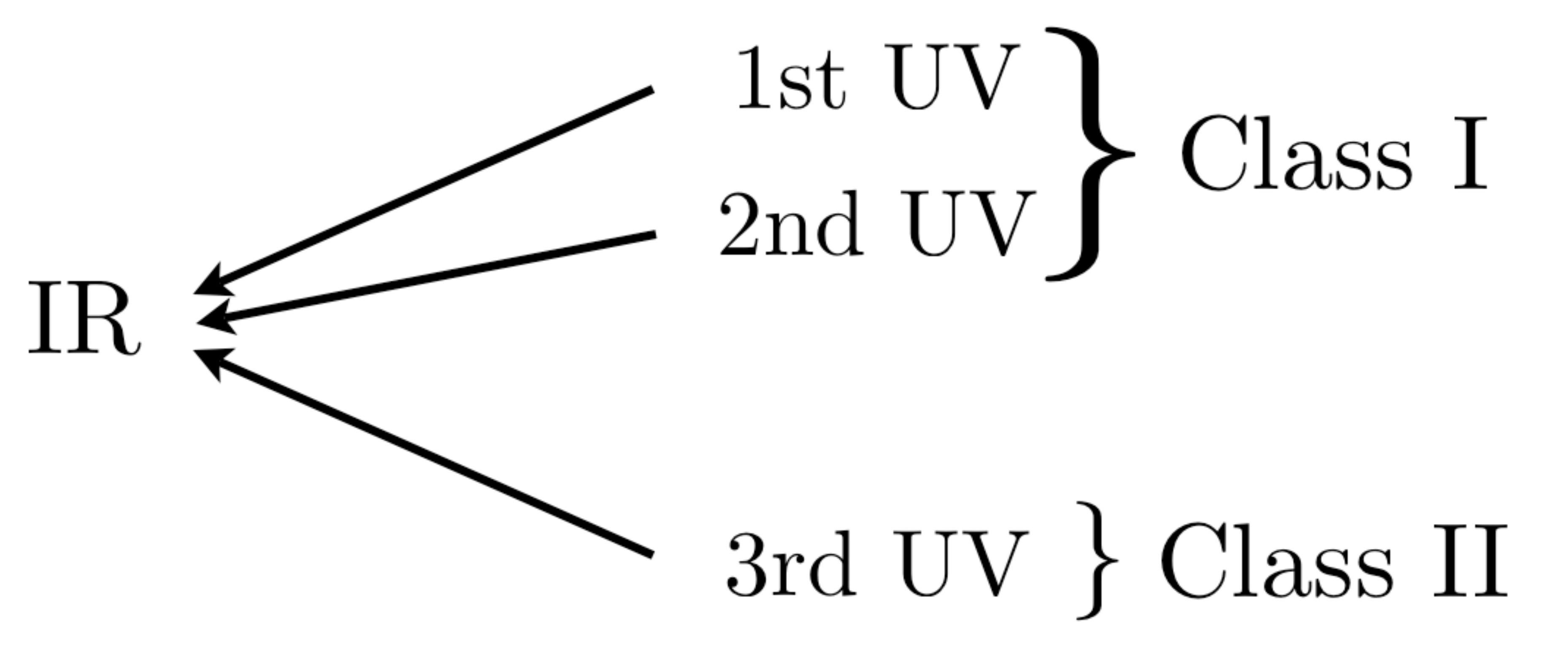}
\caption{On this picture is represented schematically the classification of
solutions in the $\HxH$ case and their RG flow.}
\label{fig:RGflow}
\end{figure}

In the gravity solutions presented in Section \ref{sec:HyperSol}, we have one IR
expansion but two possible classes of UV asymptotics. Each possibility should
correspond to a different six-dimensional UV dynamics\footnote{Since we are dealing with wrapped
branes, as we move towards the UV, we will start to see the compact directions the
brane is wrapping, and the effective four-dimensional theory living on them will
become six-dimensional. I.e., the field theory in the IR will be ``completed'' with
the dynamics of the KK modes to a different theory in the UV. Notice that since the space has a UV singularity, one would ultimately need to use string theory operations to make the theory UV complete.}. That is, we have
different solutions, each with the same IR behavior. This situation is once
again an example of the universality principle. Indeed, looking at the UV, we have
different theories. But if one follows their RG flow, one notices that they
all go to the same IR theory (see figure \ref{fig:RGflow}). As mentioned in Section
\ref{ssec:UVexpansions}, each expansion is valid only for a given range of
parameters, like $\tN_f$ and $\tN_c$. For example, the fact that the third UV is
valid only for $\tN_f < \tN_c$ means that the dual field theory cannot exhibit Kutasov
duality. The differences between the two classes of solutions will be clear in the
following sections, when studying some of the properties of their field theory
duals.

\subsection{Seeing Kutasov duality}
\label{ssec:Kutasov}

Our $\HxH$ solutions are describing D5-branes wrapping Riemann surfaces with genus
$g>1$. In the IR, one expects the theory on the branes to become effectively a
four-dimensional gauge theory and, as explained in the introduction, to have $(g-1)$
massless adjoint fermions. We will provide in what follows some arguments indicating
that we are indeed dealing with gauge theories with adjoint matter. Note that our solutions are not dual to Kutasov-like theories all the way to the UV, since they become eventually dual to six-dimensional field theories.

Kutasov duality \cite{Kutasov} is a generalization of Seiberg duality
\cite{Seiberg:1994pq}. It states the equivalence of two different $\cN=1$ gauge
theories in the IR. One is the ``electric theory'', with gauge group $SU(N_c)$,
$N_f$ quarks in the fundamental representation (and of course the corresponding
$N_f$ antiquarks in the antifundamental), and a chiral adjoint superfield $X$ with
a superpotential:
\beq
W(X)=\Tr\sum\limits_{l=1}^k g_l X^{l+1}\,,
\label{eqn:Kutasov.W}
\eeq
where $k$ is an integer. The other one is the ``magnetic theory''. It is similar,
having $N_f$ quarks (and $N_f$ antiquarks), and an adjoint chiral superfield $Y$,
but the gauge group is $SU(k N_f-N_c)$, and we also have $N_f^2$ mesons. Kutasov
duality gives a prescription for what the superpotential for $Y$ is (it is of the
type \eqref{eqn:Kutasov.W}), and for how to build the magnetic mesons out of the
electric quarks.

If one sets $k=1$, then \eqref{eqn:Kutasov.W} is just a mass term for $X$, implying
that $X$ can be integrated out in the IR; and one is left with usual SQCD, for which
Seiberg duality applies. The way one was able to see a geometric realization of
Seiberg duality in the CNP solution, was to notice that the BPS equations of the
supergravity system remained the same\footnote{There is a little subtlety here. In
principle Seiberg duality relates two different theories in the IR, while here it
would seem that the two theories related by Seiberg duality are the same. In fact,
the CNP solution is dual to $SU(N_c)$ SQCD with a quartic superpotential (generated
after integrating out the KK modes), and this theory is actually Seiberg self-dual
(see \cite{Strassler:2005qs} for a nice review). We expect a similar phenomenon for
Kutasov duality to happen in this case.} under the change:
\beq
N_c\to N_f-N_c\,,\qquad N_f\to N_f\,.
\eeq
Indeed, under this change, the only functions changing are $Q\to-Q\,,\s\to-\s$, and
this clearly leaves invariant the master equation \eqref{eqn:S2S3.master}. In the
ten-dimensional geometry, Seiberg duality is equivalent to a swap of the two-spheres
present in the $\SxS$ geometry.

It is easy to see that in our case, the master equation \eqref{eqn:H2H3.master}
possesses the symmetry:
\beq
\tN_c\to \tN_f-\tN_c\,,\qquad \tN_f\to \tN_f\,.
\eeq
If we take into account relations \eqref{eqn:tNc} and \eqref{eqn:tNf}, we can
rephrase this symmetry as:
\beq
N_c\to \frac{8\vol{\SL}}{\vol{\mH_2}^2}N_f-N_c\,,\qquad N_f\to N_f\,.
\eeq
Calling $k=\frac{8\vol{\SL}}{\vol{\mH_2}^2}$, we see that we get precisely the
transformation needed for Kutasov duality. Taking into account the way we perform
the quotients, we find that
\beq
        k = \frac{q}{g-1} \,,
\eeq
where $g$ is the genus and $q$ is a rational number. The details of this derivation
are in appendix \ref{app:quotients}. $k$ can be made an integer by choosing $g$ and
$q$ appropriately. 
Unfortunately, the relation between the quotienting and the generation of the $\Tr
X^{k+1}$ superpotential is not completely clear to us; we think it might be related
to the number of times the color branes wrap the hyperbolic cycle, as explained in
the appendix. The geometrical interpretation of Kutasov duality here would be the
swap of the two $\mH_2$'s (their quotients to be more precise) present in the
geometry \eqref{eqn:H2H3.metric}. Notice that this duality only makes sense when
$\tN_f>\tN_c$, and exchanges $2\tN_c-\tN_f\to\tN_f-2\tN_c$. In particular, it means
that it will take one solution with the asymptotics of the 1st UV \eqref{eqn:1stUV}
into one with the asymptotics of the 2nd UV \eqref{eqn:2ndUV}; and that it is not
possible to perform Kutasov duality on a solution with the asymptotics of class II. Moreover, performing a second duality gives back the original solution, analogous to what happens for Seiberg duality in CNP.

\subsection{UV behavior of the theory}
\label{ssec:Landau}

From the field theory side, not much is known about the UV of the theories
displaying Kutasov duality. The fact that in \eqref{eqn:Kutasov.W} $\Tr X^{k+1}$ is
an irrelevant operator puts these theories in need of a UV completion if
they are to be well-defined. Moreover, the general expectation from the NSVZ
$\b$-function is that we might come across a Landau pole. Since
\beq
\frac{\partial g_{YM,4}}{\partial
\log\m}\propto g_{YM,4}^{-3}\left(3N_c-N_{\rm{adj}}(1-\g_{\rm{adj}})-N_f(1-\g_f)\right)\,,
\eeq
where $N_{adj}$ is the number of chiral adjoints, and the $\g$'s are the anomalous
dimensions; we see that the adjoints will be generically pushing towards a Landau
pole, in the same direction as the flavors. It is not surprising then that our
solutions are always singular in the UV. Let us make a more precise
statement.

The gauge/gravity duality provides us with a way of computing the $\b$-function of a
gauge theory by examining the action of a brane probing the dual supergravity
solution. In our case, the computation is analogous to that carried out in
\cite{DiVecchia:2002ks,Bertolini:2002yr}. Take a D5-brane that extends on
$\mathcal{M}_4 \times \Sigma_2$, where $\Sigma_2$ is the two-cycle defined in
Section \ref{ssec:brane.setup}. We will also add a gauge field on the worldvolume of
this brane $F_{\mu\nu}$, only along the Minkowski directions. Let us first recall
that $\Sigma_2$ is defined as
\beq
        z_1=z_2, \;\;\; y_1=-y_2, \;\;\; \j=\p\,.
\eeq
A D5-brane wrapped on $\Sigma_2$ will have an induced metric on its worldvolume that
(in string frame) reads
\beq
        \td s_{ind}^2 = e^{\phi} \left[ \td x_{1,3}^2 + \left( e^{2h} + \frac{e^{2g}}{4}
(1-a)^2 \right) \frac{\td z_2^2 + \td y_2^2}{y_2^2} \right]\,,
\eeq
and the brane action will be
\beq
S= - T_{D5}\int \td^6 x\, e^{-\phi}\sqrt{-\det[g_{ab} + F_{ab}]} 
+ T_{D5}\int\left(C_6 +\frac{1}{2} C_2 \wedge F_2\wedge F_2\right)\,.
\eeq
The WZ term $C_2 \wedge F_2\wedge F_2$ will  give a theta term for the gauge field
since the $C_2 $ will be localized on the $\Sigma_2$ manifold. Now, we compute the
determinant and expand it to second order in the gauge field to get, looking at
that $F^2$ term,
\beq
        \begin{aligned}
                S &= -T_{D5}\int \td^6 x\, e^{-\phi} \frac{\sqrt{g_6}}{2}g^{\mu\nu}g^{\rho
\sigma}F_{\mu \rho}F_{\nu \sigma} \\
                &= -\left(T_{D5}\int_{\Sigma_2} \frac{\td z_2 \td y_2}{y_2^2} \right) \left[e^{2h}
+ \frac{e^{2g}}{4} (1-a)^2 \right]\int \td^4 x F^2\,.
        \end{aligned}
\eeq
So, from here we read the gauge coupling of the dual field theory that will be, up
to a constant,
\beq
\frac{1}{g_{YM,4}^2}\sim \left[e^{2h} + \frac{e^{2g}}{4} (1-a)^2 \right] .
\label{gaugecoupling}\eeq
If we now apply the change of functions from \eqref{eq:HyperChange}, we find
\beq
\frac{1}{g_{YM,4}^2}\sim - P e^{-\tau}\,.
\eeq
Starting from this expression, we can calculate the $\beta$-function for the inverse
of the gauge coupling:
\beq
        \beta_{\frac{1}{g_{YM,4}^2}} = \frac{dr}{d \log \mu}\left(\frac{d}{dr}
\frac{1}{g_{YM,4}^2} \right)\,.
\eeq
We are not going to take any precise expression for the relation between the radial
coordinate $r$ of the gravity solution and the energy scale $\mu$ of the dual field
theory, but different choices would lead to different renormalization schemes.
However, for consistency reasons, it has to be a monotonically increasing function.
Just looking at the derivative of the inverse of the coupling with respect to the
radial coordinate, we can see two different UV behaviors depending on the solution
from Section \ref{ssec:UVexpansions} we are considering:
\beq
        \bal
                \frac{d}{dr} \frac{1}{g_{YM,4}^2} &= -\frac{h_1 \tanh r_0}{2 \sqrt{-r}} + \cO
\left((-r)^0\right) & &\quad \text{for class I.} \\
                &= \left( \frac{b_0}{\cosh^2 r_0} - \tN_f \tanh r_0 \right) + \cO \left( (-r)^1
\right) & &\quad \text{for class II.}
        \eal
\eeq
We can then notice that, for class I UV asymptotics, the beta function goes to
infinity at $r = r_{UV} = 0$, which could indicate the presence of a Landau pole in
the field theory. Notice this happens regardless of the presence of flavors, in
accordance with the expectation that the adjoints might overshoot the $\b$-function.
The third UV on the contrary leads to a finite beta function even in the UV.
Nevertheless, we can see that increasing $N_f$ has the effect of of raising the
asymptotic value of the $\b$-function, once more agreeing with the field theory
expectation that the flavors should push towards a Landau pole.

In the discussion above, it is important to take into account the following remark:
the field theory will never be concerned with the part of the space close to $r = 0$
because of the behavior of the holographic {\it c}-function
\cite{Girardello:1998pd}. The latter is a quantity that was first found by reducing the ten-dimensional action to five dimensions. It is related to the number of degrees of freedom in the theory, which means that it must always increase when going from the IR to the UV. Another way to obtain it, as explained in \cite{Klebanov:2007ws}, is through the calculation of the holographic entanglement entropy, where it appears as a prefactor. The holographic entanglement entropy is computed as the volume of a minimal nine-manifold within a time slice of the ten-dimensional background, where one of the Minkowski spatial directions spans an interval. For computational details, it might be useful to have a look at \cite{Nishioka:2009un}. In our case, this volume is given by:
\beq
S_{ent}\sim\int dx \,e^{2\phi +2h +2g +k}\sqrt{1+ e^{2k }\left(\frac{\td r}{\td x}\right)^2}\,.
\eeq
From here, one can read the so-called {\it a}-charge, related to the factor in front of the square root as
\beq
        3A = 2\phi + 2h +2g+ k\,.
\eeq
Curiously, we have that $3A=\log\left(D/2\right)$, where $D$ was defined in the change of variables \eqref{eq:HyperChange}. Finally, the {\it c}-function is defined in terms of the {\it a}-charge as
\beq
        c = \frac{1}{(A')^3}\,.
\eeq
Here, we can first look at its behavior in the IR. It goes
as
\beq
        c =  27 (r-r_0)^3+{\cal O}\left((r-r_0)^4\right)\,.
\eeq
So one can see that it starts growing from the IR, and it is actually independent
from the number of flavors at first order. Then one can look at the behavior of the
{\it c}-function numerically. For every solution, the {\it c}-function becomes
infinite at some finite radius strictly before $r=0$ which we considered as the UV.
It means that the field theories dual to our solutions do not know about the whole
geometry, but rather only about the part between $r=r_0$ and the position where the
{\it c}-function blows up. 

\subsection{Domain walls}

We are first going to look at the possibility of having domain walls in our theory,
and study their tension. We will model a  domain wall (separating different vacua in
the dual field theory) by considering a five-brane that wraps a three-cycle inside
the internal geometry. We take this three-cycle to be
\beq
        \Sigma_3= [z_2, y_2, \psi]\,,
\eeq
and the brane also extends along $t,x_1, x_2$ among the Minkowski directions. Then the
induced metric on the D5-brane is
\beq
ds_{ind}^2= e^{\phi/2} \left[ dx_{1,2}^2 + 
\frac{e^{2g}}{4} \frac{\td z_2^2 + \td y_2^2}{y_2^2} +
\frac{e^{2k}}{4} \left( \td \psi + \frac{\td x_2}{y_2} \right)^2 \right]\,,
\eeq
and its action is
\beq
S= -\left[ T_{D5} \frac{e^{2\phi+2g+k}}{8}\int \frac{\td z_2 \td y_2 \td
\psi}{y_2^2}\right] \int \td^{2+1}x\,.
\eeq

The tension of this domain wall object is given by the value in the IR (as these
objects exist in the IR) of the function inside square brackets above:
\beq
T_{DW}= \frac{T_{D5} \vol{\SL}\, e^{2\phi_0}\cosh (2r_0)}{4}\frac{\sinh(2r-2r_0) 
(Q-P\coth(2r-2r_0))}{\sqrt{P^2-Q^2}}\,.
\label{tdw}
\eeq
Using the IR expansion from Section \ref{ssec:IRexp}, one can study the behavior of the
tension of the domain wall in the IR. It goes as
\beq
        T_{DW} \sim \frac{T_{D5} \vol{\SL}\, e^{2\phi_0}\cosh (2r_0)}{4} \left(1 + 2
(r-r_0)^2 + \cO \left( (r-r_0)^3 \right) \right)\,.
\eeq
So the tension of the domain wall goes to a non-zero constant in the IR. The
presence of an IR singularity casts some doubts on the validity of this result. If
we believe the fact that a good IR singularity does not spoil the physical meaning of this
computation, the result would mean that our theory has isolated vacua. It is
interesting to notice that the IR behavior of the domain wall tension does not
depend on the number of flavors $N_f$. The reason for the existence of isolated
vacua in our field theory is less obvious than in the spherical case of CNP, where
it was interpreted as a breaking of the translation invariance along $\psi$. In our
case, the function $a$ goes to a non-zero constant in the UV, so this translation
invariance does not strictly exist even in the UV of our theory. But, as the
constant towards which $a$ is going can be taken as small as one wants by moving
$r_0$ more and more towards $-\infty$, the translation invariance along $\psi$ is
still present approximately. Thus it is understandable that the domain walls behave
in the same way in both the spherical and the hyperbolic cases.

\subsection{Wilson loops}
\label{ssec:WLoops}

Another observable of the dual QFT that should be captured by our geometry is the
Wilson loop. Wilson loops provide information about the long-distance behavior of the field theory, whether it is confining, screening, etc. Through the gauge/gravity correspondence it will give us some insight about the IR geometry. 

In a gauge theory, from the expectation value of the Wilson loop in a
particular configuration, it is possible to extract the quark-antiquark potential.
The standard lore \cite{Maldacena:1998im} is that this expectation value can be
computed from the area of a certain fundamental string in the supergravity dual to
the gauge theory. The idea is to introduce a probe flavor brane (non-compact and
spanning Minkowski space-time) sitting at some $r=r_Q$ ($r_Q \sim m_Q$ is related to
the mass of the test quarks). We attach a string to this brane, whose ends
correspond to the quark and the antiquark, that will hang into the ten-dimensional
geometry, reaching a minimum radial distance $\hat{r}_0$. We can then compute the
energy $E$ of the flux-tube between the quarks as the renormalized area of the
string worldsheet, and the separation $L$ of the quarks at the end-points of the
string (measured in the Minkowski space-time) for different $\hat{r}_0$'s. We
briefly summarize the relevant formulae. For details one can have a look at
\cite{Nunez:2009da} (see also \cite{Wloops} for related examples). 

Define:
\beq
\hat{f}^2=g_{tt}g_{x^ix^i}=e^{2\f}\,,\qquad
\hat{g}^2=g_{tt}g_{rr}=e^{2\f+2k}\,,\qquad V=\frac{\hat{f}}{C
\hat{g}}\sqrt{\hat{f}^2-C^2}\,,
\label{eqn:fgV}
\eeq
where $C=\hat{f}(\hat{r}_0)$ and we are using string frame. Then,
\beq
L=2\int_{\hat{r}_0}^{r_Q}\frac{\td r}{V}\,,\qquad E=2\int_{\hat{r}_0}^{r_Q}\td
r\frac{\hat{g}\,\hat{f}}{\sqrt{\hat{f}^2-C^2}}-2\int_0^{r_Q}\td r\,\hat{g}\,.
\label{eqn:L&E}
\eeq
Several comments are in order.
First, note that the formulae in \eqref{eqn:L&E} depend on $r_Q$, that can be
interpreted as a UV regulator. Ideally one would like to take $r_Q\to\infty$, so
that the test quarks are infinitely massive and become non-dynamical. However, since
our solution never reaches infinity, we
can set at most $r_Q=r_{UV}$. Actually, as shown in Section \ref{ssec:Landau}, the
connection with the dual QFT finishes before $r=r_{UV}$. Nevertheless, one
expects the long-distance behavior of the Wilson loop to be independent of any UV
cut-off. We will critically analyze this claim in what follows.

Second, from the supergravity point of view, attaching a string to the probe flavor
brane we are introducing can be done whenever it is possible to impose Dirichlet
conditions on the string end-points. Notice that this condition is somehow also
accounting for the stability of the configuration, since it guarantees that we
can locate a flavor brane at $r=r_Q$, regardless of supersymmetry considerations.
For the type of ansatz of our geometry, as discussed in \cite{Nunez:2009da}, this is
only possible when $\displaystyle\lim_{r\to r_{Q}}V(r)=\infty$. Notice that this
only happens when the asymptotics are those of the second class of UV's (comprised
just by the so-called third UV). However, since at $r_{UV}$ we have a singularity,
it is not clear that for $r_Q\to r_{UV}$ this condition is very trustworthy, so we
will drop it when performing the numerical computations and analyze the ensuing
results.

There are two clearly differentiated regimes in which we can compute the Wilson
loop. One is the unflavored background, and the other is such that $N_f\neq0$. The
field theory expectations are different, and thus we analyze them separately:

\paragraph{$N_f=0$ geometry}

\begin{figure}
\centering
\includegraphics[width=0.5\textwidth]{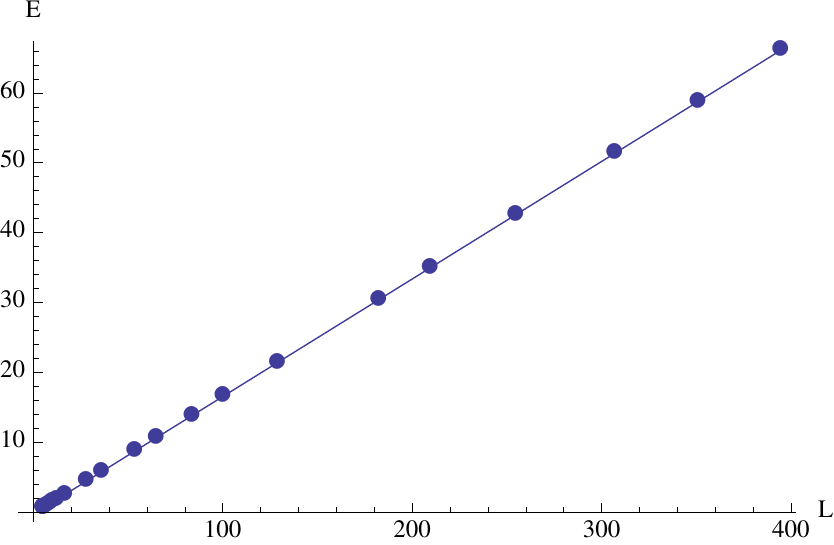}
\caption{Plot of the energy $E$ of the Wilson loop, as a function of the quark
separation $L$. We can see a linear confining behavior. This plot corresponds to a
solution with 2nd UV asymptotics, with $\tN_c = 1$ and $\tN_f = 0$.}
\label{fig:unflavor2ndUV}
\end{figure}

\begin{figure}
\centering
\includegraphics[width=\textwidth]{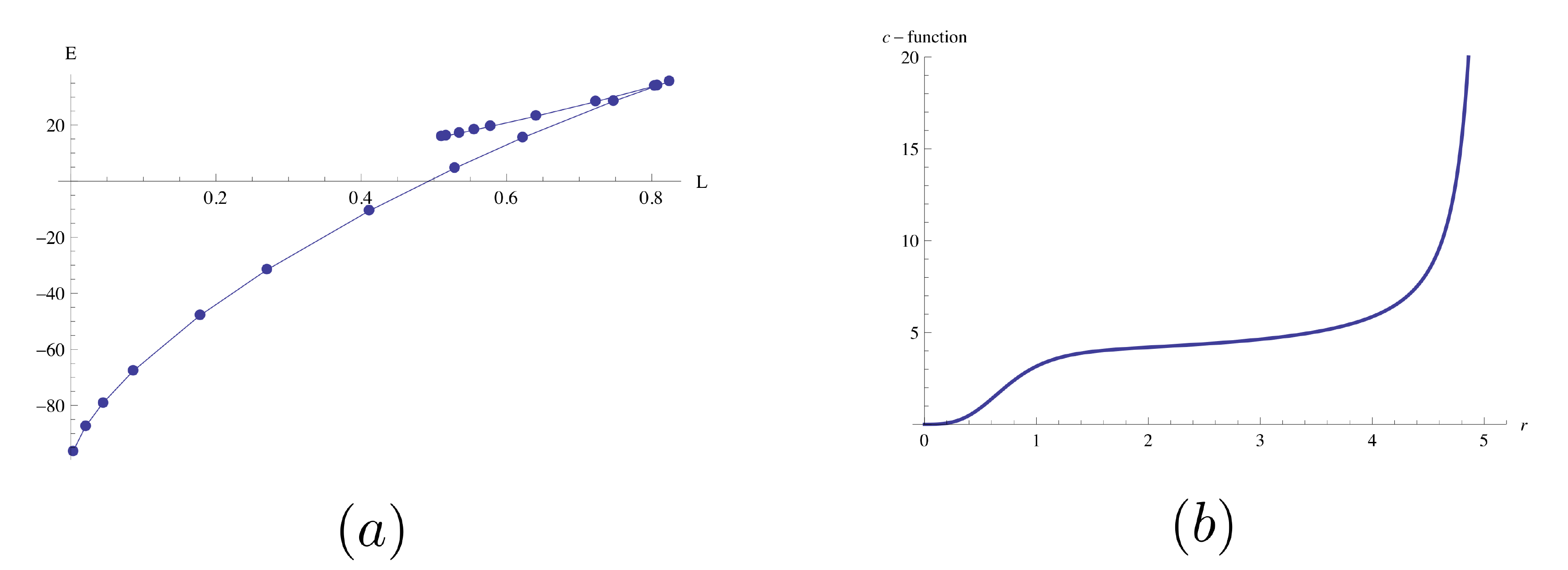}
\caption{In (a) we plot $E$ vs. $L$ for a solution with third UV asymptotics. In (b)
we plot the corresponding {\it c}-function for this solution. Notice the plateau it
shows. Both figures are with $\tN_c=1$ and $\tN_f=0$.}
\label{fig:unflavor3rdUV}
\end{figure}

The results are plotted in figure \ref{fig:unflavor2ndUV} for the asymptotics of the
first class of UV's (necessarily the second UV type, since $\tN_f=0$) and in figure
\ref{fig:unflavor3rdUV} (a) for the class II UV. There is a striking difference
between the two, since at first sight, one displays confinement, and the other one
does not. This difference is spurious though, as we now argue.

Recall the discussion in Section \ref{ssec:RG}. As we move towards the UV, the wrapped compact directions of the D5-branes will not be invisible anymore, and the gauge theory living on the stack will become six-dimensional. The different UV asymptotics we have found should be
related to different UV dynamics of this six-dimensional gauge theory. Although one would not
expect the details of the UV of the theory to affect its IR properties
from a field theory point of view, our supergravity computation of the Wilson loop
is quite sensitive to these UV details; imagine this six-dimensional dynamics is not negligible anymore from
some scale on, given by $r_{\textrm{split}}$ with $r_{IR}<r_{\textrm{split}}<r_{UV}$.
In the plots of figures \ref{fig:unflavor2ndUV} and \ref{fig:unflavor3rdUV} (a), we
take $r_{\textrm{split}}\ll r_{Q}\approx r_{UV}$, so the string giving the
Wilson loop is probing a large region of the geometry concerned with this UV dynamics,
thus rendering the results UV dependent.

The way to get rid of this issue is to shift $r_Q$ so that $r_Q\lesssim r_{\textrm{split}}$. One problem of this is that the test quarks will become dynamical.
In addition, we do not know in practice how to determine the value of $r_{\textrm{split}}$.
We think that it might be possible to estimate its value looking at figure
\ref{fig:unflavor3rdUV} (b): the fact that the {\it c}-function shows a plateau
might be signaling that, at the beginning of it, something is changing in the dual
field theory. If we identify this point with the point where the six-dimensional UV dynamics is
taking over, we have a definition for $r_{\textrm{split}}$. Performing the
numerical integration taking $r_Q=r_{\textrm{split}}$, it appears that we recover in
class II the linear confining behavior in the quark-antiquark potential, observed in
class I.

However, once the effects of the six-dimensional UV dynamics are separated, we still need to
perform a more thorough analysis of the deep IR. We have not been able to reach this
region with our numerical integration, which requires high computational precision.
This would not be very useful nonetheless: as we approach the IR singularity
$\hat{r_0}\to r_0$, from the asymptotics \eqref{eqn:HxH.IR}, it follows that $V$
would behave as $V\sim(r-r_0)^{-1/2}$. As proved in \cite{Nunez:2009da},
this will imply that the hanging string will develop a cusp near the singularity,
making the corresponding results unreliable.

As an aside note, we can also notice that, as the distance between the quarks tends
to zero, we observe a Coulombic behavior in figure \ref{fig:unflavor3rdUV} (a), but
not in figure \ref{fig:unflavor2ndUV}. We believe this is intimately related to the
discussion above about Dirichlet boundary conditions. When it is not possible to
impose those conditions, the string end-points might not be representing quarks, and
then the universal Coulombic behavior is not necessarily observed. This remark is only useful
from a pure supergravity point of view, since the connection with the
four-dimensional field theory is finishing much before the region contributing to
this effect, $r_{UV}-\eps<r<r_{UV}$.

\paragraph{$N_f\neq0$ geometry}

\begin{figure}
\centering
\includegraphics[width=\textwidth]{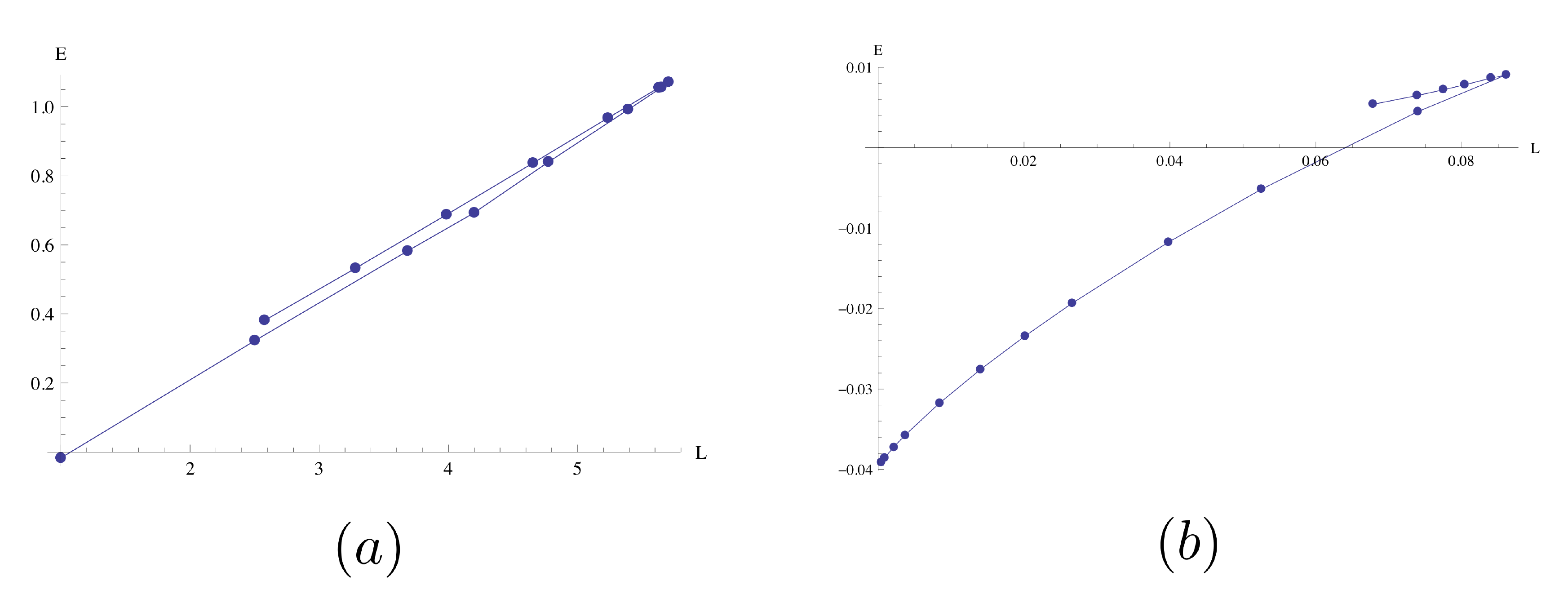}
\caption{The plot in (a) corresponds to the class I of UV asymptotics, while the one
in (b) corresponds to the UV of class II; both for $\tN_c=2$ and $\tN_f=1$.}
\label{fig:flavored}
\end{figure}

We give an example of the typical behavior for flavored solutions in figure
\ref{fig:flavored}, where plots corresponding to both classes of asymptotics have
been gathered. Let us remember that for $\tN_f>\tN_c$ we do not have a class II
solution though. In this flavored case, we expect a string-breaking length related
to the string breaking into the lightest mesons by pair production, which we observe
both in figures \ref{fig:flavored} (a) and (b). For a discussion of the effect of
smearing on this breaking length, see \cite{Bigazzi:2008gd}. Most of the comments
made in the previous $N_f=0$ case apply here. So we might expect again that these
plots are ``contaminated'' by the six-dimensional UV dynamics. Here, unfortunately, we have not
found a way to decouple this effect, since the {\it c}-function is monotonically
increasing, not showing any plateau. So the previous result and the corresponding
interpretation should be taken with a grain of salt.

Moreover, recall that the flavored solutions also have an IR singularity. According
to the criterion proposed in \cite{Maldacena:2000mw}, the singularity is good in the
weak formulation of the criterion (since $g_{tt}\sim e^{\f/2}$ is bounded), but it
is bad in its strong formulation, since the dilaton starts increasing as we move
very closely towards the singularity (see \eqref{eqn:HxH.IR}). Looking at equations
\eqref{eqn:fgV}-\eqref{eqn:L&E}, we see that the string will not be able to probe
this region in which the dilaton increases towards the singularity, indicating the
presence of some kind of IR wall. Unfortunately, this is a very delicate effect and
our numerics are not precise enough to see it.

Let us emphasize the main lessons we can draw from this section. Contrary to the usual computation, the test quarks we use are dynamical because our space does not extend to infinity. The results we obtain indicate a confining behavior of the unflavored theory. In the case with flavors, we observe the expected string-breaking phenomenon due to the pair production of quarks. However, no definite conclusion can be made due to the presence of the IR singularity.

\medskip

In the next section we leave the treatment of the $\HxH$ case to explore other possible internal spaces. The results that we will present in this following section are independent from what we have done so far. For a summary of the results of the $\HxH$ case, one can directly jump to the conclusions.

\section{The $\mH_2 \times \mS^3$ and $\mS^2 \times \SL$ ans\"{a}tze}
\label{sec:mixed}

In Section \ref{sec:HxH.ansatz}, one considered a class of metrics of the form \eqref{eqn:S2S3.metric} with the forms $\us_i$ and $\uw_i$ obeying
\beq \label{eq:HHforms}
	\bal
		&\td\uw_1=-\uw_2\wedge \uw_3\,,\quad\td\uw_2=-\uw_3\wedge \uw_1\,,\quad\td\uw_3=+\uw_1\wedge \uw_2\,. \\
		&\td\us_1=-\us_2\wedge \us_3\,,\quad\td\us_2=-\us_3\wedge \us_1\,,\quad\td\us_3=+\us_1\wedge \us_2\,.
	\eal
\eeq
And we saw that the resolution of the whole system of equations of motion of supergravity could be reduced to solving one single second-order equation, the master equation \eqref{eqn:H2H3.master}.
But instead of changing the Maurer-Cartan relations for both $\us_i$ and $\uw_i$, one can think about altering them for only one of the sets of forms. This leads to two new ans\"atze, that we are calling mixed cases. We preserve the same form for the metric, three-form and dilaton as in \eqref{eqn:H2H3.metric}-\eqref{eqn:H2H3.dil}. But we are going to take
\beq \label{eq:SHforms}
	\bal
		&\td\uw_1=-\uw_2\wedge \uw_3\,,\quad\td\uw_2=-\uw_3\wedge \uw_1\,,\quad\td\uw_3=\c \uw_1\wedge \uw_2\,. \\
		&\td\us_1=-\us_2\wedge \us_3\,,\quad\td\us_2=-\us_3\wedge \us_1\,,\quad\td\us_3=- \c \us_1\wedge \us_2\,.
	\eal
\eeq
where $\c^2 = 1$. If $\c = 1$, then we are in the case of having $\mS^2 \times \SL$, while if $\c = -1$ we have $\mH_2 \times \mS^3$. Although we preserve the same functional form for $F_3$ as in \eqref{eqn:H2H3.F3}, we will generically\footnote{Notice that the relation between for instance $\hN_c$ and $N_c$ will not be the same in the $\mS^2 \times \SL$ and the $\mH_2 \times \mS^3$ cases. We just use a common notation for convenience.} denote the parameters $\tN_c\to\hN_c$ and $\tN_f\to\hN_f$, since their proportionality relation with $N_c$ and $N_f$ respectively, will be different than in the $\HxH$ case. We can apply the same treatment as in Section \ref{sec:SUSYanalysis}, but this time we define our change of functions as
\beq\bal
	e^{2h} &= \frac{\c}{4} \frac{P^2 - Q^2}{P \cos \t - Q}\,,\qquad& a &= \frac{P \sin \t }{P \cos \t - Q}\,,\\
	e^{2g} &= \c \left(-P \cos \t + Q\right)\,,&b &= \frac{\s}{\hN_c}\,,\\
	e^{2k} &= 4 Y\,,& e^{2 \phi} &= \frac{D}{Y^{1/2}(Q^2 -P^2)} \,,\\
\eal\eeq
One can, as previously, write the BPS system of differential equations in these cases in terms of the newly defined functions:
\beq
	\bal
		&P' = 8Y - \hN_f\,,  &  &\s = \tan \t \left( Q + \frac{2\hN_c - \hN_f}{2} \right)\,, \\
		&\t' + 2 \sin \t = 0\,, & &\frac{\td}{\td r} \log \left( \frac{D}{\sqrt{Y}} \right) = \frac{16 Y P}{P^2 - Q^2}\,, \\
		&\left( \frac{Q}{\cos \t} \right)' = \frac{2\hN_c - \hN_f}{\cos^2 \t}\,, \qquad& &\frac{\td}{\td r} \log \left( \frac{D}{\sqrt{Q^2 -P^2}} \right) = 2 \cos \t \,.
	\eal
\eeq
This system is quite similar to the one of Section \ref{sec:SUSYanalysis}, and it can be solved as follows:
\beq\bal
&\s=\tan\t\,\left(Q+\frac{2\hN_c-\hN_f}{2}\right)\,,\\
&\sin\t\,=\frac{1}{\cosh(2r-2r_0)}\,,\\
&D=e^{2\f_0}\sqrt{Q^2-P^2}\cosh (2r_0)\,\cosh(2r-2r_0)\,,\\
&Y=\frac{1}{8}\left(P'+\hN_f\right)\,,\\
&Q=\left(Q_0+\frac{2\hN_c-\hN_f}{2}\right)\tanh(2r-2r_0)+\frac{2\hN_c-\hN_f}{2}\left(2r\tanh(2r-2r_0)\,-1\right)\,.
\eal\eeq
And we are left with a second-order differential equation:
\beq
	P'' + (P' + \hN_f) \left( \frac{P' + Q' + 2 \hN_f}{P - Q} + \frac{P' - Q' + 2 \hN_f}{P + Q} - 4 \tanh(2r-2r_0)\,\right) = 0\,.
\label{eq:Mixed.master}
\eeq
So in the end, in the mixed cases as well, the whole problem reduces to finding solutions of the second-order differential equation \eqref{eq:Mixed.master}. However, not all solutions of \eqref{eq:Mixed.master} are going to be valid. Indeed, they need to obey some consistency relations:
\beq
	\bal
		P' &\geq - \hN_f \,, \quad & \c \, Q &\geq \c \, P \cos \t\,, \quad  & Q^2 &\geq P^2\,.
	\eal
	\label{eqn:mixed.constraints}
\eeq
The second condition depends on $\c$, so it means that solutions valid for $\mS^2 \times \SL$ will not be valid for $\mH_2 \times \mS^3$, and vice-versa.

\subsection{Exact solutions}

In this section we present different exact solutions to the two mixed cases introduced above. Each case has been shown to reduce to the study of the same master equation \eqref{eq:Mixed.master} for the function $P(r)$. Note however, that the conditions that $P$ has to verify for each case are different, forbidding applying one solution directly to a different case.

Exact solutions extending all the way to infinity seem to exist only in the particular case where $\hN_f = 2 \hN_c$. Under this assumption, we found two exact solutions for each of the mixed cases: one that can be defined on the whole real line and that we will call type A solution by analogy with the analysis carried out in \cite{HoyosBadajoz:2008fw}; and another that starts at a finite value $r_0$ of the radial coordinate, that we will call type N solution accordingly. We also present another exact solution already known in the literature \cite{Paredes:2004xw,Radu:2002za}, that is valid only for the $\mH_2\times\mS^3$ case and $\hN_f=0$.

\subsubsection{type A solutions}
\label{ssec:typeA}

If we first look at the case where $r_0 \rightarrow - \infty$ (that is, we allow the radial coordinate $r$ to take any value in $\mathbb{R}$), we realize that the master equation reduces to:
\beq
	P'' +(P'+\hN_f) \left( \frac{P'+Q'+2\hN_f}{P-Q} + \frac{P'-Q'+2\hN_f}{P+Q} - 4 \right) = 0\,,
\eeq
where
\beq
	Q = Q_0\,.
\eeq

The solutions we found for both mixed cases can be cast in the following way:
\beq
	\bal
		P &= \hN_c - \sqrt{\hN_c^2 + Q_0^2}\,, \quad&\quad Q_0 &= 4\gamma\, \hN_c \frac{2 + \x}{\x (4+\x)}\,,
	\eal
\eeq
where $\x$ is a strictly positive constant. When $\c = 1$, then the solution corresponds to the case of having $\mS^2 \times \SL$, while if $\c = -1$ the solution is valid for the $\mH_2 \times \mS^3$ case. This translates in terms of the functions in the background as:
\beq
	\bal
		e^{2h} &= \frac{\hN_c}{\x+2(1+\gamma)}\,, \quad&\quad e^{2g} &= \frac{4 \hN_c}{\x+2(1-\gamma)}\,, \\
		e^{2k} &= \hN_c\,,\quad&\quad e^{4(\f - \f_0)} &= \frac{\x (4 + \x)}{4 \hN_c^3} e^{4r}\,.
	\eal
\eeq
These solutions are exact solutions defined for $- \infty < r < \infty$. They are singular in the IR with $e^{\f} \rightarrow 0$. And in the UV, we have $e^{\f} \rightarrow \infty$. Despite the $e^{\phi} \rightarrow 0$ singularity in the IR, this is a ``good" singularity according to the criterion of \cite{Maldacena:2000mw}. That is, the term $g_{tt}$ in the Einstein frame metric is bounded and decreasing when approaching the IR. One could then use those solutions to learn about the IR of their potential field theory dual. Let us now look at solutions of type N, where $r_0$ is finite.

\subsubsection{type N solutions}

In the following, we write two exact solutions for the case where $r_0 > - \infty$, one for each mixed case. Recall we are taking $\hN_f = 2 \hN_c$, which seems to be the only scenario where exact solutions with a good UV exist.

The solutions of the master equation read:
\beq
	\bal
		P &= -\hN_c \tanh (2r-2r_0)\,, \quad &\quad Q_0 &=\c \sqrt{3} \hN_c\,.
	\eal
\eeq
Notice that taking the limit $r_0 \rightarrow -\infty$ gives a particular solution of the type A mentioned previously.
Once again the correspondence is $\c=1$ with $\mS^2 \times \SL$, and $\c=-1$ with $\mH_2 \times \mS^3$. This means that the functions in the metric are
\beq
	\bal
		e^{2h} &= \c\frac{\hN_c}{2} \frac{\tanh (2r-2r_0)}{\tanh (2r-2r_0)+\c\sqrt{3}}\,, \\
		e^{2g} &=\c \hN_c \tanh \left(2r-2r_0\right)\left(\tanh\left(2r-2r_0\right)+\c\sqrt{3}\right)\,, \\
		e^{2k} &= \hN_c \tanh^2 (2r-2r_0)\,, \\
		e^{4\f - 4\f_0} &= \frac{2 \cosh^2 (2r -2r_0) \coth^4 (2r-2r_0)}{\hN_c^3 \cosh^2 (2r_0)}\,, \\
		a &=\frac{1}{\sinh (2r-2r_0)+\c\sqrt{3} \cosh (2r-2r_0)}\,.
	\eal
\eeq
These solutions are exact solutions defined for $r_0 < r < + \infty$. However the dilaton goes to infinity both in the IR ($r \rightarrow r_0$) and in the UV. This time the former is a bad singularity for both mixed cases. So, despite being exact, which is never easy to find, these solutions are of very little interest for our purpose, since they cannot have a well-defined field theory dual.

Dropping the $\hN_f = 2 \hN_c$ simplification, the following exact solution can also be found.

\subsubsection{A solution without flavors}

Putting $\hN_f=0$ in the master equation, one can find the following exact solution:
\beq
	\bal
		P &= 2\hN_c\, r\,, \quad &\quad Q_0 &=-\hN_c\,.
	\eal
\eeq
Looking at the constraints \eqref{eqn:mixed.constraints}, it is obvious that this solution will only work for the $\mH_2\times\mS^3$ case. Moreover, these constraints also imply that this solution will terminate at a finite value of $r$. In terms of the functions in the metric, the solution reads
\beq
	\bal
		e^{2h} &= \frac{\hN_c}{4}-\hN_c\,r\frac{2r+\sinh(4r-4r_0)}{2\cosh^2(2r-2r_0)}\,, \\
		e^{2g} &=\hN_c=e^{2k}\,, \\
		e^{4\f - 4\f_0} &= \frac{8 \cosh^4 (2r -2r_0) \cosh^2 (2r_0)}{\hN_c^3\left(1-8r^2+\cosh(4r-4r_0)-4r\sinh(4r-4r_0)\right)}\\
		&=\frac{\cosh^2(2r_0)}{\hN_c^2}\cosh^2(2r-2r_0)e^{-2h}\,, \\
		a &=\frac{2r}{\cosh (2r-2r_0)}\,.
	\eal
\eeq
This solution will be defined in an interval $\left[r_{IR},r_{UV}\right]$. It is fully regular in the IR, and it ends at $r=r_{UV}$, where $e^{2h}=0$ and the dilaton blows up.

\subsection{Asymptotic expansions in the IR}
\label{ssec:asymp.IR}

We found, for both mixed cases, four possibilities for the IR behavior of the solutions. We arranged them so that the first three expansions of each case all have a dilaton that diverges in the IR. That creates a bad singularity, that is: none of those solutions will be gravity duals to a field theory in the IR. Fortunately, the last expansion is better behaved in both cases. All of the expansions stop before reaching $r_0$, that is: $r_{IR} > r_0$.

In this section we are only going to present the different expansions for the function $P$. The corresponding results for the functions $g$, $h$, $k$, $a$ of the metric and the dilaton $\phi$ can be found in appendix \ref{app:mixed}. There, the  origin of the conditions imposed on the integration constants will be clear.
\subsubsection{$\mS^2 \times \SL$ case}

We are looking at expansions around $r_{IR}$, that is such that $r_{IR} > r_0$. Without loss of generality, we choose $r_{IR} = 0$. So the following expansions are around 0 and for $r>0$, and we have $r_0 <0$.

\paragraph{First IR}

Let us first look at the function $Q$. We parameterize its expansion as follows for convenience:
\beq
	Q = b_0 + (b_1-\hN_f) r + b_2 r^2 + \cO \left( r^3 \right)\,,
	\label{eqn:Q1stIR}
\eeq
where
\beq
	\bal
		b_0 &= \frac{1}{2} \left( \hN_f - 2\hN_c + (\hN_f - 2 \hN_c - 2 Q_0) \tanh (2r_0) \right)\,, \\
		b_1 &= \frac{1}{2 \cosh^2 (2r_0)} \left( 4 \hN_c - \hN_f + 4 Q_0 + \hN_f \cosh (4r_0) + (\hN_f- 2\hN_c) \sinh (4r_0) \right)\,, \\
		b_2 &= \frac{2}{\cosh^3 (2r_0)} \left( (2\hN_c - \hN_f) \cosh (2r_0) + (2\hN_c - \hN_f + 2Q_0) \sinh (2r_0) \right)\,.
	\eal
	\label{eqn:bs1stIR}
\eeq
As before, we first solve for the function $P$. Its expansion is
\beq \label{eq:IR11}
	P = b_0 - \hN_f r + P_2 r^2 + P_2 \frac{b_1^2 - 2 b_1 \hN_f + 2 b_0 (b_2 + 3 P_2 - 4 b_1 \tanh (2r_0))}{3b_0 b_1} r^3 + \cO \left( r^4 \right)\,,
\eeq
where $P_2$ is an integration constant that has to be taken positive.
This solution exists only in the case of $b_0 > 0$ and $b_1>0$, which corresponds to having
\beq
	Q_0 > \frac{e^{4r_0}}{1-e^{4r_0}} (2\hN_c - \hN_f)\,.
\eeq

\paragraph{Second IR}

In this paragraph, we study a second possible behavior of the functions in the IR. Once again we start with the function $Q$. The expansion is parameterized differently from before, again for convenience. Notice however that the function Q is the same:
\beq
	Q = b_0 + (b_1+\hN_f) r + b_2 r^2 + \cO \left( r^3 \right)\,,
	\label{eqn:Q2ndIR}
\eeq
where
\beq
	\bal
		b_0 &= \frac{1}{2} \left( \hN_f - 2\hN_c + (\hN_f - 2 \hN_c - 2 Q_0) \tanh (2r_0) \right)\,, \\
		b_1 &= \frac{1}{2 \cosh^2 (2r_0)} \left( 4 \hN_c - 3 \hN_f + 4 Q_0 - \hN_f \cosh (4r_0) + (\hN_f- 2\hN_c) \sinh (4r_0) \right)\,, \\
		b_2 &= \frac{2}{\cosh^3 (2r_0)} \left( (2\hN_c - \hN_f) \cosh (2r_0) + (2\hN_c - \hN_f + 2Q_0) \sinh (2r_0) \right)\,.
	\eal
	\label{eqn:bs2ndIR}
\eeq
Then we solve for the function $P$. Its expansion is
\beq \label{eq:IR12}
	P = -b_0 - \hN_f r + P_2 r^2 + P_2 \frac{b_1^2 + 2 b_1 \hN_f + 2 b_0 (b_2 - 3 P_2 - 4 b_1 \tanh (2r_0))}{3b_0 b_1} r^3 + \cO \left( r^4 \right)\,,
\eeq
where $P_2$ is an integration constant that has to be taken positive.
This solution exists only in the case of $b_0 > 0$ and $b_1>0$, which corresponds to having
\beq
	\bal
		&Q_0 > \frac{e^{4r_0}}{1-e^{4r_0}} (2\hN_c - \hN_f) & &\text{when} \quad \hN_c< \hN_f < \hN_c ( 1+e^{4r_0})\,, \\
		&Q_0 > \frac{\hN_f (3 + \cosh(4r_0))- 4 \hN_c + (2\hN_c - \hN_f) \sinh(4r_0)}{4}  & &\text{when} \quad \hN_c( 1+e^{4r_0}) \leq \hN_f < 2\hN_c\,.
	\eal
\eeq

\paragraph{Third IR}

Let us now look at another possible IR behavior. We use the same expansion for $Q$ as in the second IR discussion. Looking at $P$ we see
\beq \label{eq:IR13}
	P = -b_0 +(b_1- \hN_f) r + \frac{b_1^2 + b_1 \hN_f + b_0 (b_2 - 4 b_1 \tanh (2r_0))}{3b_0} r^2 + P_3 r^3 + \cO \left( r^4 \right)\,,
\eeq
where $P_3$ is an integration constant.
This solution exists only in the case of $b_0 > 0$ and $b_1>0$, that is, for the same ranges of values for $Q_0$ as in the second IR case.

\paragraph{Fourth IR}

In this paragraph we study the last possible IR behavior. We use the same expansion for $Q$ as in the second and third IR discussions. But this time we change our ansatz for the function $P$ by taking
\beq \label{eq:IR14}
	\bal
		P = &-b_0 + P_1 r^{1/2} + \frac{6 b_0(b_1 - \hN_f) + P_1^2}{6b_0} r\, \\
		&+ P_1 \frac{6b_0(5b_1 + 3\hN_f) + 5 P_1^2 - 72 b_0^2 \tanh(2r_0)}{72 b_0^2} r^{3/2} + \cO \left( r^2 \right)\,,
	\eal
\eeq
where $P_1$ is necessarily a positive integration constant.
This solution exists only in the case of $b_0 > 0$, which corresponds to
\beq
	Q_0 > - \frac{1}{2} (2\hN_c-\hN_f) (1+ \coth(2r_0))\,.
\eeq

\subsubsection{$\mH_2 \times \mS^3$ case}

We are now going to focus on the $\mH_2 \times \mS^3$ case. As we said, we also have four different IR expansions around $r_{IR}=0$, for $r>0$ and with $r_0 <0$. Here we compile just the different expansions for the function $P$.

\paragraph{First IR}

We choose to expand $Q$ as in \eqref{eqn:Q1stIR}-\eqref{eqn:bs1stIR}. Solving for the function $P$, we find that its expansion is naturally given by \eqref{eq:IR11} (recall the master equation is the same for both mixed cases). What changes are the conditions we have to impose on the integration constants. The integration constant $P_2$ has to be taken positive. The corresponding solution exists only in the case of $b_0 < 0$ and $b_1 < 0$, which in this case amounts to having
\beq
	\bal
		&Q_0 < \frac{e^{4r_0}}{1-e^{4r_0}} (\hN_f - 2\hN_c) & &\text{when} \quad \hN_c ( 1+e^{-4r_0}) < \hN_f\,, \\
		&Q_0 < \frac{1}{4} \left( \hN_f - 4 \hN_c - \hN_f \cosh (4r_0) + (2\hN_c - \hN_f) \sinh(4r_0) \right)  & &\text{when} \quad \hN_c( 1+e^{-4r_0}) \geq \hN_f\,.
	\eal
\eeq

\paragraph{Second IR}

If we choose to expand $Q$ as in \eqref{eqn:Q2ndIR}-\eqref{eqn:bs2ndIR}, we find a second possible behavior of the functions in the IR. Of course the function $P$ will be given by \eqref{eq:IR12}. Here $P_2$ has to be taken positive, and this solution exists only in the case of $b_0 < 0$ and $b_1 < 0$, which corresponds to having
\beq
	Q_0 < \frac{e^{4r_0}}{1-e^{4r_0}} (\hN_f - 2\hN_c)\,.
\eeq

\paragraph{Third IR}

Let us now look at another possible IR behavior. We use the same expansion for $Q$ as in the first IR. Looking at $P$ we find that
\beq \label{eq:IR23}
	P = b_0 -(b_1+ \hN_f) r - \frac{b_1^2 - b_1 \hN_f + b_0 (b_2 - 4 b_1 \tanh (2r_0))}{3b_0} r^2 + P_3 r^3 + \cO \left( r^4 \right)\,,
\eeq
where $P_3$ is an integration constant.
This solution exists only in the case of $b_0 < 0$ and $b_1<0$, that is: for the same ranges of values for $Q_0$ as in the first IR case.

\paragraph{Fourth IR}

In this paragraph we study the last possible IR behavior. We use the same expansion for $Q$ as in the first and third IRs. We will take for the function $P$ the following ansatz:
\beq \label{eq:IR24}
	P = b_0 + P_1 r^{1/2} - \frac{6 b_0(b_1 + \hN_f) + P_1^2}{6b_0} r + P_1 \frac{6b_0(5b_1 - 3\hN_f) + 5 P_1^2 - 72 b_0^2 \tanh(2r_0)}{72 b_0^2} r^{3/2} + \cO \left( r^2 \right)\,,
\eeq
where $P_1$ is necessarily a positive integration constant.
This solution exists only in the case of $b_0 < 0$, which corresponds to
\beq
	Q_0 < - \frac{1}{2} (2\hN_c-\hN_f) (1+ \coth(2r_0))\,.
\eeq

\subsection{Asymptotic expansions in the UV}
\label{ssec:asymp.UV}

The solutions for the mixed cases can reach infinity, so we only focus on these good UV's, i.e., those reaching the region $r \rightarrow \infty$. The results for both mixed cases are quite close to each other. Contrary to what happened in the $\HxH$ case, we have this time one good UV, and only one. However, it is present only in the case where $\hN_c < \hN_f < 2 \hN_c$ for the $\mS^2 \times \SL$ case, and when $\hN_f > 2 \hN_c$ for the other mixed case $\mH_2 \times \mS^3$.

As for the IR expansions, we gather in this section just the expansions for $P$. To get a better feeling of these solutions, one should check in appendix \ref{app:mixed} the asymptotic behavior of the functions $g$, $h$, $k$, $a$ of the metric and the dilaton $\phi$.

\subsubsection{$\mS^2 \times \SL$ case}

We are now going to present the UV behavior of the system. First we look at the functions $P$ and $Q$. We find the following expansions, valid for $r \rightarrow \infty$:
\beq \label{eq:UV1}
	\bal
		Q &= (2\hN_c - \hN_f) r + Q_0 + \cO \left( r^{-1} \right)\,, \\
		P &= - Q + (\hN_f - \hN_c) \left(1 - \frac{\hN_f}{4Q} + \hN_f \frac{2\hN_c - \hN_f}{8Q^2} - \hN_f \frac{16\hN_c^2 -13 \hN_c \hN_f + 2\hN_f^2}{32 Q^3} \right) + \cO \left( Q^{-4} \right)\,.
	\eal
\eeq

\subsubsection{$\mH_2 \times \mS^3$ case}

We deal with this other case in a similar fashion as above. The UV asymptotics we found for the functions $P$ and $Q$, valid for $r \rightarrow \infty$, is:
\beq \label{eq:UV2}
	\bal
		Q &= (2\hN_c - \hN_f) r + Q_0 + \cO \left( r^{-1} \right)\,, \\
		P &= Q + \hN_c \left(1 + \frac{\hN_f}{4Q} + \frac{\hN_f (\hN_f - 2\hN_c)}{8Q^2} + \frac{\hN_f(16\hN_c^2 -19 \hN_c \hN_f + 5\hN_f^2)}{32 Q^3} \right) + \cO \left( Q^{-4} \right)\,.
	\eal
\eeq

\subsection{Some comments on the solutions}

Notice that in the mixed cases we have two quantities $\hN_c$, and $\hN_f$, which should be proportional to $N_c$ and $N_f$ respectively. We have never mentioned what the relation is. Although one could expect to have relations like \eqref{eqn:tNc} and \eqref{eqn:tNf}, the reason for not having written them down is that we are not sure of what the brane setup is in these mixed cases. In both the $\SxS$ and $\HxH$ cases, the color branes were wrapping a cycle that mixed some coordinates of the two $\mS^2$'s or $\mH_2$'s present in the geometries. If this general feature holds, it is not clear to us how to entangle the coordinates of an $\mS^2$ and an $\mH_2$. As a consequence, not exactly knowing what $\hN_c$ and $\hN_f$ stand for, we have not pursued a further analysis of the connection of these solutions with their field theory duals. In any case, we would like to make a couple of remarks about them.

A good place to start a possible investigation of the field theory could be the solutions of Section \ref{ssec:typeA}. Indeed, they are analytic well-behaved solutions, very similar to the one found in Section 6 in CNP. The fact that the metric functions are constant might make it easier to find a way to compute the gauge coupling, even if the cycle on which to wrap a probe D5-brane is not clear. In the aforementioned solution of CNP, the gauge coupling is constant, in accordance to the field theory expectation of a ``conformal point'' $2N_c=N_f$. One could try to learn about the field theory dual to our solution looking for an analogue of this fact.

Regarding the UV, the solutions for these cases are better behaved than their non-mixed relatives, for it is possible for some of them to reach infinity, at least for some combinations of $\hN_c$ and $\hN_f$. In the IR, they can always flow to a geometry with a singularity of the good type. It is also an interesting fact that exact solutions could be found in this case, at least in the case $\hN_f = 2 \hN_c$. When considering the unflavored setup $\hN_f = 0$, other exact solutions exist, at least in the $\mH_2\times\mS^3$ case; this can be somewhat related to the fact that one can uplift on an $\mS^3$ a $D=7$ $SO(4)$ gauged supergravity solution (see Section 7.2.3 of \cite{Paredes:2004xw}). However, this solution does not go all the way to infinity in the UV. We are not aware of a $D=7$ gauged supergravity generated by the compactification of type IIB supergravity on a Bianchi group (for M-theory this construction was done in \cite{Bergshoeff:2003ri}).

Finally, one can wonder if, due to the presence of hyperbolic cycles in the geometries, there is some Kutasov-like duality here. In principle, the transformation $Q\to-Q\,,\s\to-\s$ leaves the master equation invariant. However, if we look at the solutions with nice IR and UV behaviors, we see that the Kutasov-like duality $\hN_c\to\hN_f-\hN_c\,,\,\hN_f\to\hN_f$ will interchange a solution of the $\mH_2\times\mS^3$ case with one of the $\mS^2\times\SL$ case, and vice-versa. This complies with the geometrical interpretation of this Kutasov-like duality as a swap in a given geometry of the $\mH_2$ and the $\mS^2$. The fact that we are lacking the correct identification of $\hN_c$ and $\hN_f$ in these mixed cases makes the field-theoretical interpretation of this fact far from obvious.

\section{Conclusions}

In this paper, we looked into the possibility of finding gravity duals to field
theories exhibiting a Kutasov-like duality. The existence of chiral adjoint
superfields in the field theory was ensured by having branes wrapped around cycles
of higher genus on the gravity side. That is why we studied supergravity solutions
where the internal space contains hyperbolic subspaces that, once properly
quotiented, will have submanifolds of non-trivial homology. More precisely, we investigated
three possible types of internal manifolds containing a fibered  product of either
$\mH_2 \times \SL$, $\mS^2 \times \SL$ or $\mH_2 \times \mS^3$. We showed that the
search for solutions in each case could be reduced to solving a ``master" equation,
that is only one second-order ordinary differential equation for a function $P$
obeying some constraints. For the first case, despite the fact that the master
equation was the same as in previously studied cases, we found that it was not
possible to get solutions going all the way to infinity in the UV. The end of the space introduces a singularity in the supergravity solution; that was expected from the field theory which needs a UV completion.  We presented
several asymptotic solutions. The solutions are
singular in the IR, but it is always a good singularity. For the mixed
cases, we found several exact and asymptotic solutions. In the case $N_f\neq0$, all of them have good UV's, but they are singular in the IR.

In Section \ref{sec:FT}, we presented some features of the field theories
dual to the solutions of the background $\mH_2 \times \SL$. After discussing the way
our different supergravity solutions are related through RG flow, we looked at how
Kutasov duality is implemented by a quotienting of the hyperbolic spaces by
subgroups. We showed that, depending on how these subgroups are chosen, the $k$
parameter of the Kutasov duality can take different values. We studied the gauge
coupling of the theory in the UV, matching some qualitative expectations from the
field theory, as well as the holographic {\it c}-function, discovering that one
needs to put a UV cut-off before the end of the space, due to the divergence of the
{\it c}-function at a finite point in the radial direction. We also investigated the
domain walls and the Wilson loops. Those calculations are not fully reliable because
of the IR singularity. The domain wall tension, which does not depend at first order
on the number of flavors, indicates the existence of isolated vacua. Concerning the Wilson loop calculation, the presence of the UV singularity forced us to use dynamical test quarks. The results are different for the flavored and the unflavored solutions. For $N_f=0$, we obtain indications of confinement. For $N_f \neq 0$, the flux-tube between the quarks can decay into mesons, which is reflected in the string-breaking phenomenon we observe.

It would be interesting for the future to get more details concerning the field
theories dual to our solutions. In particular, to get a better handle on some loose
comments we made along the way, like for instance the way the superpotential for the
adjoints is generated on the supergravity side. The $N_f=0$ solution has a
particularly nice IR singularity, and it could be interesting to investigate further
details of it: e.g., the origin of the plateau the {\it c}-function was displaying
in this case. In addition, we have not said anything on the possible duals to the
solutions of the mixed cases. In those cases it is indeed much less clear how to
translate quantities from the gravity picture to the field theory. Even finding the
cycle wrapped by the branes is far from obvious. Finally, it could be interesting to
specify a given quotient of the hyperbolic spaces in order to study the dual field
theory in more detail.

\section*{Acknowledgements}

We are very grateful to Carlos N\'u\~nez and Alfonso V. Ramallo for their continuous
encouragement and guidance. We would like to thank also Daniel Are\'an, Paolo
Benincasa, Aldo Cotrone, Chethan Krishnan, Joyce Myers, Jos\'e Antonio Oubi\~na, \'Angel Paredes
and Johannes Schmude for useful discussions and comments on the manuscript. E.C.
would like to thank Swansea University for their hospitality when this project was
started. This work was supported in part by MICINN and  FEDER  under grant
FPA2008-01838,  by the Spanish Consolider-Ingenio 2010 Programme CPAN
(CSD2007-00042) and by Xunta de Galicia (Conselleria de Educacion  grant INCITE09
206 121 PR). E.C. is supported by a Spanish FPU fellowship. E.C. thanks the FRont Of
Galician-speaking Scientists for unconditional support.

\appendix

\section{A different hyperbolic parameterization}

We compile here an alternative explicit realization of the set of one-forms used in \eqref{eqn:us's} and \eqref{eqn:uw's} to characterize $\mH_2$ and $\SL$ respectively. They mimic the ones used in  \cite{Casero:2006pt} when studying the $\SxS$ case.
\beq
\us_1=\td z_1\,,\quad\us_2=\sinh z_1\,\td y_1\,,\quad\us_3=\cosh z_1\,\td y_1\,,
\label{eqn:us's.2}
\eeq
would be the ones for $\mH_2$, where we are using $\{z_1,y_1\}$ as its coordinates, and the metric reads as before: $\td s^2=\left(\us_1\right)^2+\left(\us_2\right)^2$. Using coordinates $\{z_2,y_2,\z\}$ for $\SL$, a possible set of left-invariant forms is:
\beq\begin{aligned}
\uw_1&=\frac{1}{\sqrt{\sinh^2\z+\cosh^2\z}}\left(\cosh \z\,\td z_2+\sinh \z\sinh z_2\, \td y_2\right)\,,\\
\uw_2&=\frac{1}{\sqrt{\sinh^2\z+\cosh^2\z}}\left(-\sinh \z\,\td z_2+\cosh \z\sinh z_2\, \td y_2\right)\,,\\
\uw_3&=\frac{1}{\sqrt{\sinh^2\z+\cosh^2\z}}\td \z+\cosh z_2\,\td y_2\,.
\label{eqn:uw's.2}
\end{aligned}\eeq
The metric of $\SL$ would still read as in \eqref{eqn:metric.SL2}. Notice that if we think of $\SL$ as a line bundle over $\mH_2$, we could identify the base $\mH_2$ as that spanned by $\{z_2,y_2\}$ and define a new fiber coordinate by:
\beq
\frac{1}{\sqrt{\cosh2\z}}\td \z=\td \j\quad\Rightarrow\quad\tanh \z=\tan\j\,,
\eeq
which in turn implies
\beq
\frac{\cosh \z}{\sqrt{\cosh2\z}}=\cos\j\,,\qquad\frac{\sinh \z}{\sqrt{\cosh2\z}}=\sin\j\,.
\eeq
This allows us to identify this coordinate $\j$ with the one used in \eqref{eqn:uw's}. Notice that choosing for $\z$ its maximum range, $-\infty<\z<\infty$, it seems we only cover half of the range $\j$ has ($0\leq\j<2\p$ since it is parameterizing the complex line). This suggests that the coordinates $\{z_2,y_2,\z\}$ may be just parameterizing $PSL(2,\mathbb{R})$, which is known to be double-covered by $\SL$.

\section{UV problem of the $\mH_2 \times \SL$ case}
\label{sec:badUV}

We will prove here that any solution of the master equation for the $\HxH$ case \eqref{eqn:H2H3.master} will always break down at some finite value of $r$. Recall that in order for the solutions to be consistent we need the following conditions to hold:
\beq
P\leq 0\,,\qquad\qquad |Q|\leq|P|\,,\qquad\qquad P'+\tN_f\geq 0\,.
\label{eqn:mEconditions}
\eeq
Let us proceed by contradiction:

\medskip

Assuming we have a solution extending all the way from some finite $r_{\textrm{IR}}$ to $\infty$, if we look at the conditions \eqref{eqn:mEconditions} for large enough $r$, we easily deduce that
\beq
-\tN_f\,\leq\,\displaystyle\lim_{r\to\infty}P'\,\leq\,0\,.
\label{eqn:limit}
\eeq
Now let us focus our attention on the $r\to\infty$ limit of the following piece of the master equation:
\beq
\frac{P' + Q' + 2 \tN_f}{P - Q} + \frac{P' - Q' + 2 \tN_f}{P + Q}\,.
\label{eqn:piece}
\eeq
We want to see that the limit of this piece is not positive. When $2\tN_c=\tN_f$, which implies that $Q$ is constant, it is immediate that this limit is negative or zero in virtue of the constraints \eqref{eqn:mEconditions}. In the $2\tN_c\neq\tN_f$ case, we can notice that these constraints imply that both denominators are always negative, and also that the $P'+\tN_f$ piece is always positive. Since asymptotically we have $Q'+\tN_f\sim2\tN_c$, the first summand will give a non-positive contribution. The second summand is a little bit more troublesome, since $-Q'+\tN_f\sim2(\tN_f-\tN_c)$ asymptotically, and this could be negative if $\tN_f>\tN_c$. But actually, when $\tN_f>\tN_c$ holds, one can see that because of the last constraint in \eqref{eqn:mEconditions}, the denominator $P+Q$ goes to $-\infty$, and the contribution of this summand is null.

So we conclude that the $r\to\infty$ limit of \eqref{eqn:piece} is not positive. We can then have a look at the limit of the whole master equation \eqref{eqn:H2H3.master}:

Assuming that $P$ is monotonic for large $\r$, which is a sensible physical condition to impose, one can rigorously prove that \eqref{eqn:limit} implies $\displaystyle\lim_{r\to\infty}P''=0\;$. Then:
\beq\bal
0=\lim_{r\to\infty}P''&=-\lim_{r\to\infty}\left[(P'+ \tN_f)\left(\frac{P'+Q'+2\tN_f}{P-Q}+\frac{P'-Q'+2\tN_f}{P+Q}-4\coth(2r-2r_0)\right)\right]\leq\\
&\leq-4(\lim_{r\to\infty}P'+\tN_f)\,.
\eal\eeq
The only possibility for satisfying this equation is to have $\displaystyle\lim_{r\to\infty}P'=-\tN_f$. But actually this is ruled out by the master equation as well. This can be seen by writing $P=-\tN_f r+p(r)$, with $p(r)$ tending to zero as $r\to\infty$. The master equation could be solved asymptotically and the leading behavior for $p$ would be $p\sim e^{4r}$: a contradiction.
\bigskip

So the assumption that a solution of the master equation satisfying the constraints \eqref{eqn:mEconditions} would exist all the way till $r\to\infty$ leads us to a contradiction. Thus, any solution of the master equation fulfilling our requirements will eventually break down.

\section{How to quotient $\mH_2$ and $\SL$}
\label{app:quotients}

We briefly discuss in this appendix what are the possible quotients by discrete groups of isometries we can perform on $\mH_2$ and $\SL$, and what is the resulting value for the ratio
\beq
k=\frac{8\vol{\SL}}{\vol{\mH_2}^2}\,,
\label{eqn:k}
\eeq
which we have associated in Section \ref{ssec:Kutasov} with the integer number appearing in \eqref{eqn:Kutasov.W}, relevant for Kutasov duality. Recall that in \eqref{eqn:k}, the volumes stand for the finite volumes of the quotients $\mH_2/\Gamma$ and $\SL/G$.

The quotients of $\mH_2$ are very well known. The discrete subgroups $\Gamma$ of its isometry group $PSL(2,\mathbb{R})$ are the so-called Fuchsian groups, and the resulting quotients $\mH_2/\Gamma$ are Riemann surfaces of genus $g>1$ of constant negative curvature $R=-1$. The volume of such a quotient can be straightforwardly computed from the Gauss-Bonnet theorem:
\beq
\vol{\mH_2}=\int\o_{\vol{\mH_2}}=-\int R\,\o_{\vol{\mH_2}}=-2\p\chi(g)=4\p(g-1)\,,
\label{eqn:vol.H2}
\eeq
where $\chi$ is the Euler characteristic of the resulting Riemann surface.

The isometry group of $\SL$ might be less well known, but its structure can be deduced from the exact sequence
\beq
0\to\mathbb{R}\to\mathcal{I}\to PSL(2,\mathbb{R})\to1\,,
\eeq
where $\mathcal{I}$ is standing for the identity component\footnote{The isometry group of $\SL$ has two connected components and the other one simply contains the isometries induced from the orientation-reversing isometries of $\mH_2$.} of the isometry group of $\SL$. This means that basically there are two types of isometries acting on $\SL$, that recall can be thought as an $\mS^1$ bundle over $\mH_2$. One type comprises the isometries that rotate the $\mS^1$, i.e., that rotate the fibers through a constant angle, while covering the identity map of $\mH_2$. This type is parameterized by $\mathbb{R}$. The other type is composed of those isometries that ``rotate'' the base $\mH_2$, and it is therefore parameterized by $PSL(2,\mathbb{R})$. This ``rotation'' on the base also induces a constant-angle rotation in each fiber $\mS^1$.

The idea to retain from the discussion of the paragraph above, is that each quotient of $\SL$ will be roughly a quotient of the base $\mH_2$ times a quotient of $\mS^1$. The quotient we have to perform in the base $\mH_2$ has to be equal to the one we performed in the other $\mH_2$ of the geometry. The only freedom left is to perform an extra discrete quotient in $\mS^1$. We compute the volume of $\SL$, including the effect of a winding number $m$ of the color branes, as:
\beq
\vol{\SL}=m\int\o_{\vol{\SL}}=m\int\o_{\vol{\mH_2}}\,\o_{\vol{\mS^1}}=m\vol{\mH_2}\vol{\mS^1}\,.
\eeq
We already know the volume of the base \eqref{eqn:vol.H2}. The volume of the $\mS^1$, taking into account the quotienting, will be $\vol{\mS^1}=\frac{2\p}{n}$, where $n$ is an integer. Then:
\beq
\vol{\SL}=2\p^2q\,(g-1)\,,
\eeq
where $q=\frac{4m}{n}$ is a rational number. And coming back to \eqref{eqn:k}, the $k$ of Kutasov duality will be, in terms of the quotient parameters:
\beq
k=\frac{q}{g-1}\,.
\label{eqn:final.k}
\eeq
In general $q\in\mathbb{Q}$, but for some particular configurations, this $k$ will become an integer. As we see, $k$ is proportional to the winding number $m$ of the color branes wrapping the hyperbolic cycle. We think this might be the reason $k$ is appearing in the superpotential for the adjoint fermions in the dual field theory: as an adjoint can be thought as a zero-mode of the B-field wrapping a particular cycle on the Riemann surface, the winding of the brane would correspond to the adjoint self-interacting $k\sim m$ times.

\section{Details of the solutions for the mixed ans\"{a}tze}
\label{app:mixed}

In this appendix we collect the details of the expansions for the functions in the metric and the dilaton, for each solution found in Section \ref{sec:mixed}. 

\subsection{Asymptotic expansions in the IR}

These expansions complement Section \ref{ssec:asymp.IR}.

\subsubsection{$\mS^2 \times \SL$ case}

\paragraph{First IR}

Using the expansion from \eqref{eq:IR11}, one finds
{\allowdisplaybreaks\begin{align}
	e^{2h} = &\frac{b_1}{2 + 2 \tanh(2r_0)} r \nonumber\\
	&+ \frac{-b_1^2 + 2 b_0 (2 b_1 + b_2 - P_2) + (b_1^2 + 2 b_0 (b_2 - P_2)) \tanh(2r_0) - 4 b_0 b_1 \tanh^2 (2r_0)}{4 b_0 (1+ \tanh(2r_0))^2} r^2 \nonumber\\
	&+ \cO \left( r^3 \right) \,,\nonumber\\
	e^{2g} = &b_0 \left(1 + \tanh(2r_0) \right) + \left(b_1 + 2 b_0(\tanh^2 (2r_0) - 1) - N_f (1+\tanh(2r_0)) \right) r \nonumber\\
	&+ \left( b_2 + P_2 \tanh(2r_0)+(2 N_f - 4 b_0 \tanh(2r_0)) (1- \tanh^2 (2r_0)) \right) r^2 + \cO \left(r^3 \right)\,, \nonumber\\
	e^{2k} = &P_2 r + P_2 \frac{b_1^2 - 2 b_1 N_f + 2 b_0 (b_2 + 3 P_2 - 4 b_1 \tanh (2r_0))}{2b_0 b_1} r^2 + \cO \left(r^3 \right) \,,\nonumber\\
	a = &\left( \sinh(2r_0) - \cosh(2r_0) \right) + \frac{b_1 - 2 b_0 \left(1 + \tanh(2r_0) \right)}{b_0 \cosh(2r_0) \left(1+\tanh(2r_0) \right)^2} r + \cO \left( r^2 \right)\,, \nonumber\\
	e^{4\phi} = &e^{4\phi_0} \left(\frac{2}{b_0 b_1 P_2} \frac{1}{r^2} - 2 \frac{b_1^2 - 2b_1 N_f + 2b_0 (b_2 + P_2)}{b_0^2 b_1^2 P_2} \frac{1}{r} + \cO \left(r^0 \right) \right)\,.
\end{align}}
We see that the dilaton is divergent in the IR, while $e^{2h}$ and $e^{2k}$ go to zero.

\paragraph{Second IR}

Using the expansion from \eqref{eq:IR12}, one finds
\beq
	\bal
		e^{2h} = &\frac{b_1}{2 - 2 \tanh(2r_0)} r \\
		&+ \frac{-b_1^2 + 2 b_0 (-2 b_1 + b_2 + P_2) - (b_1^2 + 2 b_0 (b_2 + P_2)) \tanh(2r_0) + 4 b_0 b_1 \tanh^2 (2r_0)}{4 b_0 (1- \tanh(2r_0))^2} r^2 \\
		&+ \cO \left(r^3 \right)\,, \\
		e^{2g} = &b_0 \left(1 - \tanh(2r_0) \right) + \left(b_1 - 2 b_0(\tanh^2 (2r_0) - 1) + N_f (1-\tanh(2r_0)) \right) r \\
		&+ \left( b_2 + P_2 \tanh(2r_0)+(2 N_f + 4 b_0 \tanh(2r_0)) (1- \tanh^2 (2r_0)) \right) r^2 + \cO \left(r^3 \right)\,, \\
		e^{2k} = &P_2 r + P_2 \frac{b_1^2 + 2 b_1 N_f + 2 b_0 (b_2 - 3 P_2 - 4 b_1 \tanh (2r_0))}{2b_0 b_1} r^2 + \cO \left(r^3 \right)\,, \\
		a = &\left( \sinh(2r_0) + \cosh(2r_0) \right) - \frac{b_1 + 2 b_0 \left(-1 + \tanh(2r_0) \right)}{b_0 \cosh(2r_0) \left(1-\tanh(2r_0) \right)^2} r + \cO \left( r^2 \right)\,, \\
		e^{4\phi} = &e^{4\phi_0} \left(\frac{2}{b_0 b_1 P_2} \frac{1}{r^2} - 2 \frac{b_1^2 + 2b_1 N_f + 2b_0 (b_2 - P_2)}{b_0^2 b_1^2 P_2} \frac{1}{r} + \cO \left(r^0 \right) \right)\,.
	\eal
\eeq
We see that as before, the dilaton is divergent in the IR, while $e^{2h}$ and $e^{2k}$ go to zero.

\paragraph{Third IR}

Using the expansion from \eqref{eq:IR13}, one finds
\beq
	\bal
		e^{2h} = &\frac{b_1}{1 - \tanh(2r_0)} r\, \\
		&+ \frac{4b_0 (b_2-3b_1)+b_1(N_f-5b_1) - (4b_0(b_1+b_2)+b_1(7b_1+N_f)) \tanh(2r_0) + 16 b_0 b_1 \tanh^2 (2r_0)}{6 b_0 (1- \tanh(2r_0))^2} r^2 \\
		&+ \cO \left(r^3 \right)\,, \\
		e^{2g} = &b_0 \left(1 - \tanh(2r_0) \right) + \left(b_1 + N_f + (b_1-N_f) \tanh(2r_0) - 2 b_0(\tanh^2 (2r_0) - 1) \right) r + \cO \left( r^2 \right)\,, \\
		e^{2k} = &\frac{b_1}{2} + \frac{b_1^2 + b_0 b_2 + b_1 N_f - 4 b_0 b_1 \tanh(2r_0)}{3b_0} r + \frac{3P_3}{2} r^2 + \cO \left(r^3 \right)\,, \\
		a = &\left( \sinh(2r_0) - \cosh(2r_0) \right) - 2 \frac{-b_1 + b_0 \left(1 + \tanh(2r_0) \right)}{b_0 \cosh(2r_0) \left(1+\tanh(2r_0) \right)^2} r + \cO \left( r^2 \right)\,, \\
		e^{4\phi} = &e^{4\phi_0} \left( \frac{2}{b_0 b_1^2} \frac{1}{r} - \frac{5 b_1^2 + 11 b_1 N_f + 8b_0 b_2}{3 b_0^2 b_1^3} + \cO \left(r \right) \right)\,.
	\eal
\eeq
We see that this time, the dilaton is divergent in the IR, while $e^{2h}$ alone goes to zero.

\paragraph{Fourth IR}

Using the expansion from \eqref{eq:IR14}, one finds
\beq
	\bal
		e^{2h} = &\frac{P_1}{2 - 2\tanh(2r_0)} r^{1/2} + \frac{-P_1^2 (1+2 \tanh(2r_0))+ 6 b_0 b_1 (1-\tanh(2r_0)) }{6 b_0 (1- \tanh(2r_0))^2} r + \cO \left(r^{3/2} \right)\,, \\
		e^{2g} = &b_0 \left(1 - \tanh(2r_0) \right) + P_1 \tanh(2r_0) r^{1/2} \\
		&+ \frac{6b_0 \left(2b_0 (1-\tanh^2(2r_0))+ b_1 + N_f + (b_1 - N_f) \tanh(2r_0) \right) + P_1^2 \tanh(2r_0)}{6b_0} r \\
		&+ \cO \left(r^{3/2} \right)\,, \\
		e^{2k} = &\frac{P_1}{4} r^{-1/2} + \frac{6b_0 b_1 + P_1^2}{12b_0} + P_1 \frac{6b_0(5b_1 + 3N_f) + 5 P_1^2 - 72 b_0^2 \tanh(2r_0)}{96 b_0^2} r^{1/2} + \cO \left(r^{3/2} \right)\,, \\
		a = &\left( \sinh(2r_0) + \cosh(2r_0) \right) -\frac{P_1}{b_0 \cosh(2r_0) (1-\tanh(2r_0))^2} r^{1/2} \\
		&+ \frac{12 b_0^2 (1-\tanh(2r_0))^2 +12 b_0 b_1 (1-\tanh(2r_0)) +P_1^2 (1-7 \tanh(2r_0))}{6 b_0^2 \cosh(2r_0) (\tanh(2r_0) - 1)^3} r + \cO \left(r^{3/2} \right)\,, \\
		e^{4\phi} = &e^{4\phi_{IR}} \left(1 - \frac{4 b_1}{P_1} r^{1/2} + \cO \left(r \right) \right)\,.
	\eal
\eeq
For this IR the dilaton is well-behaved. There is a good singularity at $r=0$, and if one wants in addition to have the dilaton decreasing towards the IR, one requires taking $b_1 < 0$. Moreover, $e^{2h}$ goes to zero while $e^{2k}$ diverges.

\subsubsection{$\mH_2 \times \mS^3$ case}

\paragraph{First IR}

Here we find
{\allowdisplaybreaks\begin{align}
		e^{2h} = &-\frac{b_1}{2 + 2 \tanh(2r_0)} r \nonumber\\
		&+ \frac{b_1^2 - 2 b_0 (2 b_1 + b_2 - P_2) - (b_1^2 + 2 b_0 (b_2 - P_2)) \tanh(2r_0) + 4 b_0 b_1 \tanh^2 (2r_0)}{4 b_0 (1+ \tanh(2r_0))^2} r^2\nonumber \\
		&+ \cO \left(r^3 \right)\,,\nonumber \\
		e^{2g} = &-b_0 \left(1 + \tanh(2r_0) \right) + \left(-b_1 + 2 b_0(1-\tanh^2 (2r_0)) + N_f (1+\tanh(2r_0)) \right) r\nonumber \\
		&- \left( b_2 + P_2 \tanh(2r_0)+(2 N_f - 4 b_0 \tanh(2r_0)) (1- \tanh^2 (2r_0)) \right) r^2 + \cO \left(r^3 \right)\,,\nonumber \\
		e^{2k} = &P_2 r + P_2 \frac{b_1^2 - 2 b_1 N_f + 2 b_0 (b_2 + 3 P_2 - 4 b_1 \tanh (2r_0))}{2b_0 b_1} r^2 + \cO \left(r^3 \right)\,,\nonumber \\
		a = &\left( \sinh(2r_0) - \cosh(2r_0) \right) + \frac{b_1 - 2 b_0 \left(1 + \tanh(2r_0) \right)}{b_0 \cosh(2r_0) \left(1+\tanh(2r_0) \right)^2} r + \cO \left( r^2 \right)\,,\nonumber\\
		e^{4\phi} = &e^{4\phi_0} \left(\frac{2}{b_0 b_1 P_2} \frac{1}{r^2} - 2 \frac{b_1^2 - 2b_1 N_f + 2b_0 (b_2 + P_2)}{b_0^2 b_1^2 P_2} \frac{1}{r} + \cO \left(r^0 \right) \right)\,.
\end{align}}
We see that as before, the dilaton is divergent in the IR, while $e^{2h}$ and $e^{2k}$ go to zero.

\paragraph{Second IR}

The expansions are:
\beq
	\bal
		e^{2h} = &\frac{b_1}{-2 + 2 \tanh(2r_0)} r\\
		&+ \frac{-b_1^2 + 2 b_0 (2 b_1 - b_2 - P_2) + (b_1^2 + 2 b_0 (b_2 + P_2)) \tanh(2r_0) - 4 b_0 b_1 \tanh^2 (2r_0)}{4 b_0 (1- \tanh(2r_0))^2} r^2 \\
		&+ \cO \left(r^3 \right)\,, \\
		e^{2g} = &b_0 \left(-1 + \tanh(2r_0) \right) - \left(b_1 + 2 b_0(1-\tanh^2 (2r_0)) + N_f (1-\tanh(2r_0)) \right) r \\
		&- \left( b_2 + P_2 \tanh(2r_0)+(2 N_f + 4 b_0 \tanh(2r_0)) (1- \tanh^2 (2r_0)) \right) r^2 + \cO \left(r^3 \right) \,,\\
		e^{2k} = &P_2 r + P_2 \frac{b_1^2 + 2 b_1 N_f + 2 b_0 (b_2 - 3 P_2 - 4 b_1 \tanh (2r_0))}{2b_0 b_1} r^2 + \cO \left(r^3 \right) \,,\\
		a = &\left( \sinh(2r_0) + \cosh(2r_0) \right) - \frac{b_1 + 2 b_0 \left(1 - \tanh(2r_0) \right)}{b_0 \cosh(2r_0) \left(1-\tanh(2r_0) \right)^2} r + \cO \left( r^2 \right)\,,\\
		e^{4\phi} = &e^{4\phi_0} \left(\frac{2}{b_0 b_1 P_2} \frac{1}{r^2} - 2 \frac{b_1^2 + 2b_1 N_f + 2b_0 (b_2 - P_2)}{b_0^2 b_1^2 P_2} \frac{1}{r} + \cO \left(r^0 \right) \right)\,.
	\eal
\eeq
We see that in that particular case, the dilaton is divergent in the IR, while $e^{2h}$ and $e^{2k}$ go to zero.

\paragraph{Third IR}

Using the expansion from \eqref{eq:IR23}, one finds
\beq
	\bal
		e^{2h} = &-\frac{b_1}{1 + \tanh(2r_0)} r \\
		&+ \frac{-4b_0 (b_2+3b_1)+b_1(N_f+5b_1) + (4b_0(b_1-b_2)+b_1(N_f-7b_1)) \tanh(2r_0) + 16 b_0 b_1 \tanh^2 (2r_0)}{6 b_0 (1+ \tanh(2r_0))^2} r^2 \\
		&+ \cO \left(r^3 \right)\,, \\
		e^{2g} = &-b_0 \left(1 + \tanh(2r_0) \right) + \left(N_f - b_1 + (N_f + b_1) \tanh(2r_0) - 2 b_0(\tanh^2 (2r_0) - 1) \right) r + \cO \left( r^2 \right)\,, \\
		e^{2k} = &-\frac{b_1}{2} - \frac{b_1^2 - b_1 N_f +b_0 (b_2 -4b_1 \tanh(2r_0))}{3b_0} r + \frac{3P_3}{2} r^2 + \cO \left(r^3 \right)\,, \\
		a = &\left( \sinh(2r_0) + \cosh(2r_0) \right) - 2 \frac{b_1 + b_0 \left(1 - \tanh(2r_0) \right)}{b_0 \cosh(2r_0) \left(1-\tanh(2r_0) \right)^2} r + \cO \left( r^2 \right)\,, \\
		e^{4\phi} = &e^{4\phi_0} \left(-\frac{2}{b_0 b_1^2} \frac{1}{r} + \frac{5 b_1^2 - 11 b_1 N_f + 4b_0 (2b_2+b_1 \tanh(2r_0))}{3 b_0^2 b_1^3} + \cO \left(r \right) \right)\,.
	\eal
\eeq
We see that this time, the dilaton is divergent in the IR, while $e^{2h}$ alone goes to zero.

\paragraph{Fourth IR}

Finally, using the expansion from \eqref{eq:IR24}, one finds
\beq
	\bal
		e^{2h} = &\frac{P_1}{2 + 2\tanh(2r_0)} r^{1/2} + \frac{P_1^2 (1-2 \tanh(2r_0)) - 6 b_0 b_1 (1+\tanh(2r_0)) }{6 b_0 (1+ \tanh(2r_0))^2} r + \cO \left(r^{3/2} \right) \,,\\
		e^{2g} = &-b_0 \left(1 + \tanh(2r_0) \right) - P_1 \tanh(2r_0) r^{1/2} \\
		&+ \frac{6b_0 \left(2b_0 (1-\tanh^2(2r_0)) +N_f - b_1 + (N_f + b_1) \tanh(2r_0) \right) + P_1^2 \tanh(2r_0)}{6b_0} r + \cO \left(r^{3/2} \right)\,, \\
		e^{2k} = &\frac{P_1}{4} r^{-1/2} - \frac{6b_0 b_1 + P_1^2}{12b_0} + P_1 \frac{6b_0(5b_1 - 3N_f) + 5 P_1^2 - 72 b_0^2 \tanh(2r_0)}{96 b_0^2} r^{1/2} + \cO \left(r^{3/2} \right)\,, \\
		a = &\left( \sinh(2r_0) - \cosh(2r_0) \right) -\frac{P_1}{b_0 \cosh(2r_0) (1+\tanh(2r_0))^2} r^{1/2} \\
		&+ \frac{-12 b_0^2 (1+\tanh(2r_0))^2 +12 b_0 b_1 (1+\tanh(2r_0)) +P_1^2 (1+7 \tanh(2r_0))}{6 b_0^2 \cosh(2r_0) (\tanh(2r_0) + 1)^3} r + \cO \left(r^{3/2} \right)\,, \\
		e^{4\phi} = &e^{4\phi_{IR}} \left( 1 + \frac{4 b_1}{P_1} r^{1/2} + \cO \left(r \right) \right)\,.
	\eal
\eeq
We see that, this time, the dilaton is well well-behaved in the IR. In addition, $e^{2h}$ goes to zero while $e^{2k}$ diverges.

\subsection{Asymptotic expansions in the UV}

The expansions that follow concern Section \ref{ssec:asymp.UV}.

\subsubsection{$\mS^2 \times \SL$ case}

Using the expansion from \eqref{eq:UV1}, one finds
\beq
	\bal
		e^{2h} = &\frac{N_f-N_c}{4} \left(1 + \frac{N_f}{4N_f - 8N_c} r^{-1} + N_f \frac{2N_c - N_f + 2Q_0}{8 (2N_c-N_f)^2} r^{-2} \right) + \cO \left( r^{-3} \right)\,, \\
		e^{2g} = &2 (2N_c-N_f) r + N_c - N_f + 2Q_0 + N_f \frac{N_f-N_c}{8N_c - 4N_f} r^{-1} + \cO \left( r^{-2} \right)\,, \\
		e^{2k} = &(N_f - N_c) \left(1 + \frac{N_f-N_c}{16 N_c - 8N_f} r^{-2} + N_f \frac{N_f - 2N_c -2Q_0}{8 (2N_c-N_f)^2} r^{-3} \right) + \cO \left( r^{-4} \right)\,, \\
		a = &e^{-2r} \left( 1 + \frac{N_c -N_f}{4N_c -2N_f} r^{-1} + \frac{(N_f-N_c)(2N_c - N_f +4Q_0)}{8 (2N_c-N_f)^2} r^{-2} + \cO \left( r^{-3} \right) \right)\,, \\
		e^{4\phi- 4\phi_0} = &e^{4r} \left( \frac{1}{2 \cosh^2(2r_0) (N_f-N_c)^2 (2N_c-N_f)} r^{-1} \right. \\
		&\qquad \left. - \frac{2N_c-3N_f + 4Q_0}{8 \cosh^2(2r_0) (2N_c^2 - 3 N_c N_f +N_f^2)^2} r^{-2} + \cO \left( r^{-3} \right) \right)\,.
	\eal
\eeq
Here we notice that the dilaton and the function $g$ diverge at infinity, while the functions $h$ and $k$ go to constant. In addition, the function $a$ goes to zero, which means that the fibration between the  $\mS^2$ and $\SL$ disappears at infinity.

\subsubsection{$\mH_2 \times \mS^3$ case}

Using the expansion from \eqref{eq:UV2}, one finds
\beq
	\bal
		e^{2h} = &\frac{N_f - 2 N_c}{2} r - \frac{N_c + 2Q_0}{4} + \frac{N_c N_f}{16(N_f-2N_c)} r^{-1} \cO \left( r^{-2} \right)\,, \\
		e^{2g} = &N_c \left( 1 + \frac{N_f}{8N_c - 4N_f} r^{-1} + N_f \frac{N_f - 2 N_c - 2Q_0}{8 (N_f-2N_c)^2} r^{-2} \right) + \cO \left( r^{-3} \right)\,, \\
		e^{2k} = &N_c \left(1 + \frac{N_f}{8N_f - 16 N_c} r^{-2} + N_f \frac{2N_c -N_f +2Q_0}{8 (N_f - 2N_c)^2} r^{-3} \right) + \cO \left( r^{-4} \right)\,, \\
		a = &e^{-2r} \left( \frac{4N_c-2N_f}{N_c} r + \frac{4N_c - N_f + 4 Q_0}{2N_c} + \frac{N_f(4N_c-N_f)}{N_c(16 N_c -8N_f)} r^{-1} + \cO \left( r^{-2} \right) \right)\,, \\
		e^{4\phi- 4\phi_0} = &e^{4r} \left( \frac{1}{2 \cosh^2(2r_0) N_c^2 (N_f-2N_c)} r^{-1} + \frac{2N_c+N_f + 4Q_0}{8 \cosh^2(2r_0) N_c^2 (N_f -2N_c)^2} r^{-2} + \cO \left( r^{-3} \right) \right)\,.
	\eal
\eeq
Here we notice that the dilaton and the functions $h$ diverge at infinity, while the functions $g$ and $k$ go to constant.  We notice that $a \rightarrow 0$ in the UV, effectively killing the fibration between $\mH_2$ and $\mS^3$.


\begin{thebibliography}{99}
\bibitem{Perelman}
  G.~Perelman,
  ``Finite extinction time for the solutions to the Ricci flow on certain
three-manifolds,''
  [math/0307245 [math-dg]].
  
  G.~Perelman,
  ``The Entropy formula for the Ricci flow and its geometric applications,''
  [math/0211159 [math-dg]].
  
  G.~Perelman,
  ``Ricci flow with surgery on three-manifolds,''
    [math/0303109 [math-dg]].
\end{thebibliography}
\end{document}